\documentclass[usenatbib]{mn2e} 
\usepackage{times}
\usepackage{amssymb} 
\usepackage{amsmath} 
\usepackage{graphicx}
\pdfoutput=1

\title[Radiative transfer in disc galaxies IV]%
{Radiative transfer in disc galaxies IV -- The effects of dust
  attenuation on bulge and disc structural parameters}

\author[D. A. Gadotti et al.]{Dimitri A. Gadotti$^{1,2}$\thanks{E-mail: dgadotti@eso.org}, Maarten Baes$^3$
  and Sarah Falony$^3$
  \\
  $^1$Max-Planck-Institut f\"ur Astrophysik, Karl-Schwarzschild-Strasse 1, D-85748 Garching bei M\"unchen, Germany \\
$^2$European Southern Observatory, Casilla 19001, Santiago 19, Chile\\  $^3$Sterrenkundig Observatorium, Universiteit Gent, Krijgslaan
  281-S9, B-9000 Gent, Belgium }

\begin{document}

\maketitle

\begin{abstract}
  Combining Monte Carlo radiative transfer simulations and accurate 2D
  bulge/disc decompositions, we present a new study to investigate the
  effects of dust attenuation on the apparent structural properties of
  the disc and bulge of spiral galaxies. We find that dust affects the
  results from such decompositions in ways which cannot be identified
  when one studies dust effects on bulge and disc components
  separately. In particular, the effects of dust in galaxies hosting pseudo-bulges
might be different from those in galaxies hosting classical bulges, even if their dust
content is identical. Confirming previous results, we find that disc scale
  lengths are overestimated when dust effects are important. In
  addition, we also find that bulge effective radii and S\'ersic
  indices are underestimated. Furthermore, the apparent attenuation of
  the integrated disc light is underestimated, whereas the
  corresponding attenuation of bulge light is overestimated. Dust
  effects are more significant for the bulge parameters, and,
  combined, they lead to a strong underestimation of the bulge-to-disc
  ratio, which can reach a factor of two in the $V$ band, even at
  relatively low galaxy inclinations and dust opacities. Nevertheless,
  it never reaches factors larger than about three, which corresponds to
a factor of two in bulge-to-total ratio. Such effect
can have an impact on studies of the black hole/bulge scaling relations.
\end{abstract}


\section{Introduction}

The vital role of interstellar dust as an important component of the
interstellar medium in galaxies has been demonstrated extensively. One
of the least popular aspects of interstellar dust, at least for those
astronomers who consider dust as a nuisance rather than a fascinating
field per se, is the efficiency at which it absorbs and
scatters UV/optical radiation. It has been known for many decades that
the presence of dust influences the observed,
apparent photometric galaxy parameters (apparent scale lengths,
surface brightnesses, luminosities, axial ratios, etc.) and makes it a
challenge to recover the intrinsic, unaffected parameters.

Several authors have investigated these effects using radiative
transfer modelling with varying degrees of sophistication and/or
geometrical realism. In general, it was found that the importance of dust
attenuation varies as a function of wavelength, galaxy inclination and
star-dust geometry \citep[e.g.][]{1988ApJ...333..673B,
  1994ApJ...432..114B, 1994MNRAS.266..511E, 1996A&A...313..377D,
  1996ApJ...465..127B, 2001MNRAS.326..733B, 2001MNRAS.323..130C}. The
most up-to-date study on this topic is the recent investigation by
\citet{2006A&A...456..941M}. These authors investigate the systematic
effects of dust attenuation on the apparent scale lengths, central
surface brightnesses and axial ratios in pure disc galaxies as a
function of inclination and dust mass. They find that dust can
significantly affect both the scale length and central surface
brightness, inducing variations in the apparent to intrinsic
quantities of up to 50\% in scale length and up to 1.5 magnitudes in central
surface brightness in the $B$ band.

Unfortunately, these studies are typically restricted to the disc
component only. Nevertheless, it is extremely important to investigate
the effects of dust attenuation on the bulge parameters as well.
Bulges hold clues to different galaxy formation and evolution
processes \citep[see e.g.][and references therein]
{2004ARA&A..42..603K, 2005MNRAS.358.1477A, 2009MNRAS.393..1531G}.
Galaxies hosting classical bulges are believed to have had
evolutionary histories with a significant contribution from mergers,
as opposed to galaxies hosting pseudo-bulges, with a more quiet
evolution.  In fact, \citet{2008MNRAS.390..881D} present results which
suggest that pseudo-bulges are related to isolated galaxies. Like
discs, bulge structural parameters are also expected to be affected by
dust attenuation. Indeed, \citet{2007MNRAS.379.1022D} find empirically
that about twice as many photons produced in bulges are absorbed by
dust as photons produced in discs.

In the last few years, two independent teams have investigated the
effects of dust attenuation in bulge and disc
components, on their integrated properties, separately, using realistic models of
spiral galaxies. \citet{2004ApJ...617.1022P} presented attenuation
functions for the individual disc and bulge components of dusty spiral
galaxies, based on the {\sc dirty} Monte Carlo radiative transfer code
\citep{2001ApJ...551..269G}. In a similar way, \citet[][see also
\citealt{2008ApJ...678L.101D}]{2004A&A...419..821T} investigated the
attenuation of the bulge and disc components in spiral galaxies using
the scattered-intensities method pioneered by
\citet{1987ApJ...317..637K} and \citet{1994ApJ...432..114B}. Both
teams clearly demonstrated that the effects of dust on the bulge and
disc components can differ substantially, as a result of the different
star-dust geometry.

In the framework of our series of papers on radiative transfer in disc
galaxies \citep{2001MNRAS.326..722B, 2001MNRAS.326..733B,
  2003MNRAS.343.1081B}, we aim to extend this approach one stage
further and want to investigate the systematic effects of dust
attenuation on the apparent detailed structural properties of discs
and bulges {\em simultaneously}. As mentioned, dust attenuation effects on
the determination of structural parameters of pure discs, and on
integrated properties of individual bulges and discs, have been studied.
However, most massive galaxies contain a bulge and a disc, and the
determination of structural parameters for the different galactic
components through 2D image decomposition has become a popular tool
\citep[e.g.][and references therein] {2005MNRAS.362.1319L,
  2006MNRAS.371....2A, 2008MNRAS.388.1708G, 2009MNRAS.393..1531G}.
How dust affects the direct output of such decompositions still needs
to be evaluated. This is the main goal of the current study. As we
will see, the complexity of such decompositions can lead to unforseen
effects. Dust effects in one galaxy component can alter the parameters
of the other component even if dust is absent in the latter.

In order to perform such study, we have created several realistic,
dusty disc galaxy models, with different dust opacities and at various
inclination angles, and carried out sophisticated 2D bulge/disc
decompositions on these models. This paper is organised as
follows. The next section describes how the models were created and
the 2D fits carried out. Section 3 presents how disc and bulge
parameters obtained through the 2D fits vary with dust opacity and
inclination angle. The results are discussed in Section 4. In Section
5 we discuss possible implications of our results on galaxy structure
studies.  Finally, in Section 6 we summarise our main conclusions.

\section{The modelling process}

\subsection{Input models}

\begin{table*}
  \centering
  \caption{Parameters describing the dusty disc galaxy models in the $H$ band.}
  \label{skirt.tab}
  \begin{tabular}{lccc} \hline\hline
  parameter & unit &          TrueBulge & PseudoBulge
 \\ \hline
  H-band disc luminosity $L_{\text{disc},H}$ & $10^9~L_\odot$ &  70.8 & 67.0 \\ 
  stellar disc scale length $h_R$ & kpc           &  4.0    & 4.0    \\ 
  stellar disc scale height $h_z$  & kpc           &  0.35 & 0.35 \\ 
  stellar disc SSP age $t_{\text{disc}}$ & Gyr          & 4.0      & 4.0 \\
  H-band bulge luminosity $L_{\text{bulge},H}$ & $10^9~L_\odot$ & 47.2  & 7.45 \\ 
  bulge S\'ersic index $n$   & ---                        & 3.5   &  1.5 \\ 
  bulge effective radius $R_{\text{e}}$ & kpc           & 1.5   &  0.75 \\
  bulge SSP age $t_{\text{bulge}}$ & Gyr           & 11.0   & 4.0  \\
  H-band total luminosity $L_{\text{tot},H}$ & $10^9~L_\odot$ & 118  & 74.5 \\
  H-band bulge fraction $(B/T)_H$  & ---                     & 0.4  & 0.1  \\
  dust disc scale length $h_{R,{\text{d}}}$ & kpc              & 4.0  & 4.0  \\
  dust disc scale height $h_{z,{\text{d}}}$ & kpc              & 0.14  & 0.14  \\
  $V$-band face-on optical depth $\tau$ & ---                   & 0\ldots8 & 0\dots8 \\ 
    \hline\hline
  \end{tabular}
\end{table*}

The input models are a subset of an extensive library of disc galaxy
models, set up to investigate in a systematic way the attenuation,
dust temperature structure and FIR/submm emission
\citep{Baes-atlas}. The models considered in this paper consist of a
double-exponential stellar disc, a spherical stellar S\'ersic bulge
and a double-exponential dust disc\footnote{By double-exponential disc
  we mean a disc whose stellar/dust surface density varies
  exponentially in both the radial and vertical directions.}. To cover
both early-type and late-type galaxies, we consider two types of
models, denoted as TrueBulge and PseudoBulge models. The models we
consider are based on the results of \citet{2004A&A...414..905H}, who
obtained the structural parameters of 108 disc galaxies using $H$-band
observations, and the work of \citet{2009MNRAS.393..1531G}, who similarly
performed 2D decomposition
of almost 1000 galaxies using images from the SDSS DR2 in the $g$, $r$ and $i$
bands. Our models thus mimic the typical galaxy hosting a classical bulge
and the typical galaxy hosting a pseudo-bulge, according to observational results.

The stellar disc has 4 free parameters: the radial scale length
$h_R$, the vertical scale length, or scale height, $h_z$, the stellar age
$t_{\text{disc}}$ (the intrinsic spectral energy distribution is
represented as a simple stellar population with solar metallicity --
hereafter SSP) and the bolometric luminosity $L_{\text{disc,bol}}$ (or
equivalently, the luminosity $L_{\text{disc},X}$ in a given band
$X$). The TrueBulge model is characterized by a disc with scale length
$h_R=4$~kpc, $H$-band luminosity $L_{\text{disc},H} =
7.08\times10^{10}~L_\odot$ and stellar age $t_{\text{disc}} =
4$~Gyr. The vertical scale length of the stellar disc was not
recovered in the studies mentioned above; based on our own Milky
Way, we adopt the value $h_z=350$~pc. This results in $h_R/h_z\sim11$,
a value in agreement with studies of edge-on spiral galaxies
\citep{1998MNRAS.299..595D, 2000A&A...361..451S}. The PseudoBulge
model disc has the same scale length, scale height and SSP age, but
the disc is slightly less luminous with
$L_{\text{disc},H}=6.70\times10^{10}~L_\odot$.

The stellar bulge is also characterized by 4 parameters: the effective
radius $R_{\text{e}}$, the S\'ersic index $n$, the age
$t_{\text{bulge}}$ and the bolometric luminosity
$L_{\text{bulge},bol}$ (or $L_{\text{bulge},X}$). For the TrueBulge
model we adopt the structural values $R_{\text{e}}=1.5$~kpc and
$n=3.5$. The stellar population is assumed to be relatively old and
red at $t_{\text{bulge}}=11$~Gyr. With a luminosity of
$L_{\text{bulge},H} = 4.72\times10^{10}~L_\odot$, the bulge
contributes 40\% of the total luminosity in the $H$ band. The bulge of
the PseudoBulge model is smaller, less concentrated, younger and
fainter; the parameters are $R_{\text{e}}=750$~pc, $n=1.5$,
$t_{\text{bulge}}=4$~Gyr and
$L_{\text{bulge},H}=7.45\times10^9~L_\odot$. In the $H$ band, this
comes down to a contribution of only 10\%.

As mentioned, the dust is also distributed in a double-exponential
disc. The geometrical parameters are similar to those adopted by
\citet{1994ApJ...432..114B} and we assume similar dust properties for
the TrueBulge and PseudoBulge models. The scale length of the dust
disc is the same as the scale length of the stellar disc
($h_R=4$~kpc). In contrast, the dust disc in spiral galaxies must be
much thinner than the stellar disc in order to be able to generate the
prominent dust lanes seen in edge-on galaxies. The scale height of the
dust disc is taken to be 40\% of the stellar disc scale height, in
agreement with detailed radiative transfer studies of edge-on spiral
galaxies \citep{1999A&A...344..868X, 2007A&A...471..765B}. For the
optical properties of the dust, we adopted the BARE\_GR\_S model from
\citet{2004ApJS..152..211Z}. This model represents a realistic dust
mixture of bare (i.e. non-composite) graphite, silicate and PAH dust
grains. The size distribution of each of these dust grain populations
is fine-tuned in such a way that the global dust properties accurately
reproduce the extinction, emission and abundance constraints of the
Milky Way. The final parameter that characterizes the dust component
in our galaxy models is the total dust mass, which is a free parameter
in our models. Following the custom in spiral galaxies radiative
transfer studies, we characterize the dust mass by the face-on optical
depth in the optical $V$ band, defined as the integral of the opacity
along the symmetry axis of the galaxy, \begin{equation}
  \tau_{\text{V}} = \int_{-\infty}^{\infty} \kappa_{\text{V}}
  \rho_{\text{d}}(0,z)\, {\text{d}}z
\end{equation}
where $\kappa_{\text{V}}$ is the extinction coefficient at the centre
of the $V$-band model and $\rho_{\text{d}}$ is the dust mass density. In the
remainder of this paper we will denote the $V$-band optical depth simply
as $\tau$. We consider optical depths $\tau=0$, 0.2, 0.5, 1, 2, 4, 6 and
8. Table {\ref{skirt.tab}} presents the relevant parameters used to
create the dusty disc galaxy models. It should be stressed that the dust
component in our models is diffuse. A clumpy distribution of dust results
in less conspicuous effects than a diffuse distribution with the same
total dust mass.

\subsection{Radiative transfer simulations}
\label{rts}

The dust-affected galaxy images are constructed with the 3D Monte
Carlo radiative transfer code {\sc skirt} \citep{2003MNRAS.343.1081B,
  2005AIPC..761...27B}. {\sc skirt} was originally developed to study
the effects of dust absorption and scattering on the observed
kinematics of galaxies. It has grown to be a flexible Monte Carlo
radiative transfer code that can be used to model a variety of dusty
systems, from spiral and elliptical galaxies to circumstellar dust
discs. It takes into account the physical processes of absorption,
multiple anisotropic scattering and thermal re-emission by dust grains.
Further, the code is strongly optimized, using many well-known and some
novel Monte Carlo radiative transfer optimization techniques, including
forced first scattering, photon peel-off, continuous absorption,
polychromatism and smart detectors \citep[see
e.g.][]{1970A&A.....9...53M, 1977ApJS...35....1W, 1984ApJ...278..186Y,
  1999A&A...344..282L, 2001ApJ...554..615B, 2005NewA...10..523B,
  2005A&A...440..531J, 2006MNRAS.372....2J, 2008MNRAS.391..617B}. The
general output of the code are global and spatially resolved SEDs and
images for arbitrary viewing points and at any wavelength from the UV
to the mm range.

Given that the calculation of the volume emissivity of the bulge component is not trivial, and the fact that our results are fundamentally based on this calculation, we describe in Appendix A how {\sc skirt} deals with it.

The model images used here are $V$-band images created from a viewing
point with inclinations $i$ of 15, 30, 45 and 60 degrees with respect to
the plane of the disc of the galaxy (a face-on projection corresponds to
$i=0^\circ$). We did not investigate galaxies
at an inclination higher than 60 degrees, as for these galaxies the
scale height of the disc plays an important and complicating role. A
circular Gaussian function is convolved with the images in order to
simulate the effects of atmospheric turbulence and instrumental
resolution. The Gaussian variance was chosen as to produce a point
spread function with full width at half maximum typical of most
imaging data from ground-based telescopes, corresponding to 75 pc in our simulated images. In total we investigated 64
images. The simulated images contain $481\times481$ pixels with a
pixel size of 50~pc, resulting in a total field of view of
$24~{\text{kpc}}\times24~{\text{kpc}}$.

\subsection{Bulge/disc decomposition}

\begin{figure*}
  \centering
  \includegraphics[viewport=35 125 575 660,clip,width=0.33\textwidth]{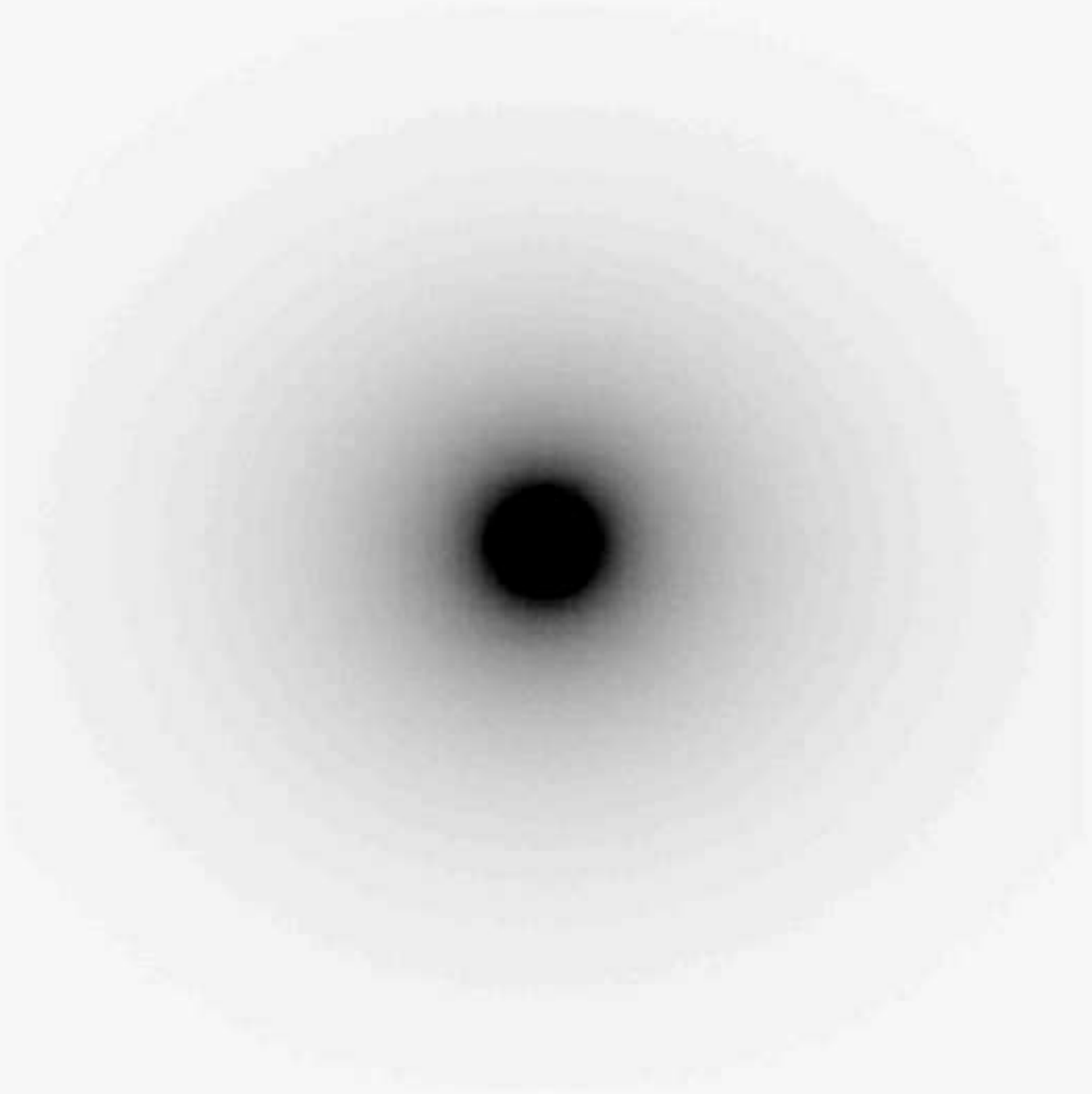}
  \includegraphics[viewport=35 125 575 660,clip,width=0.33\textwidth]{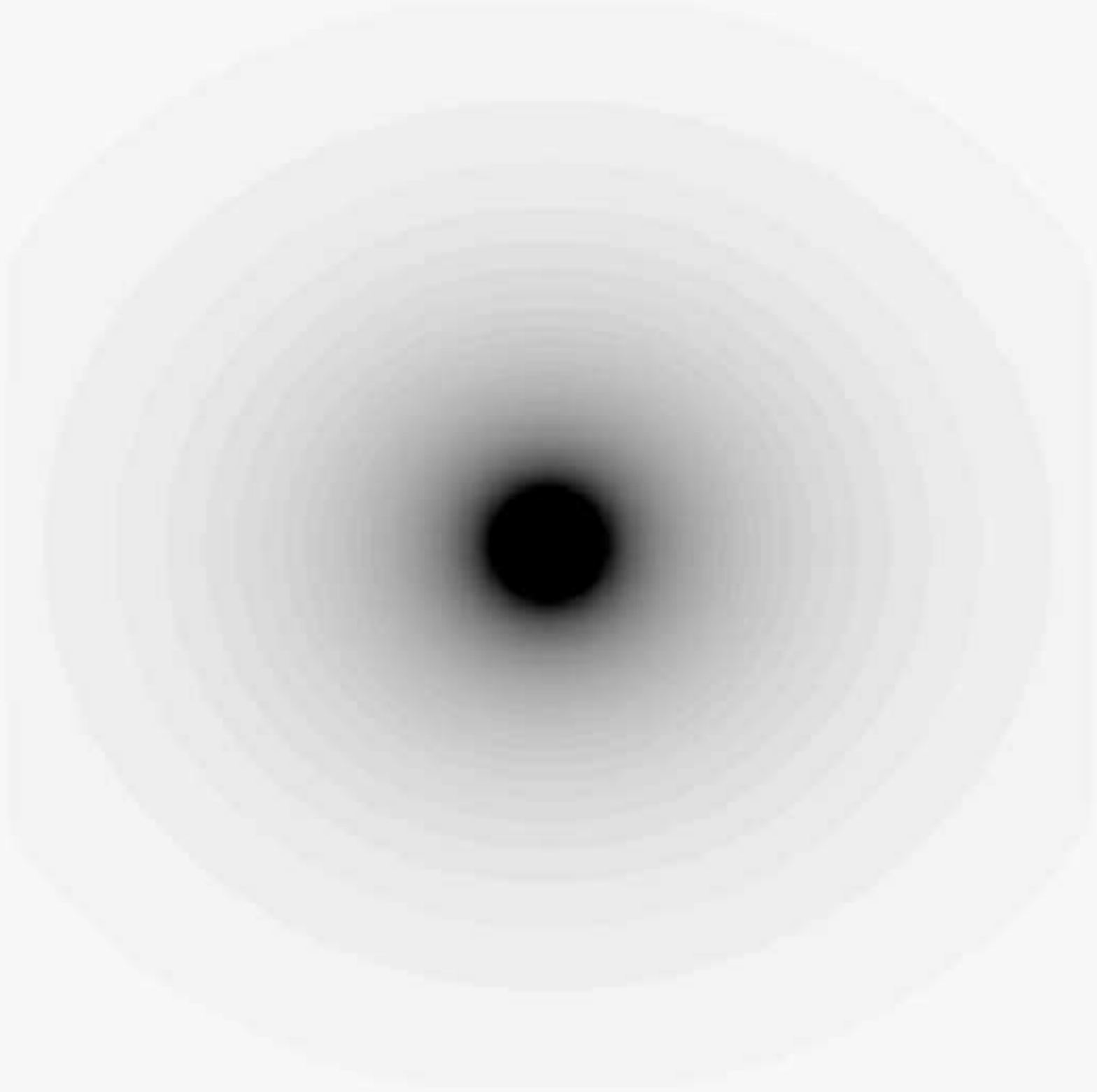}
  \includegraphics[viewport=35 125 575 660,clip,width=0.33\textwidth]{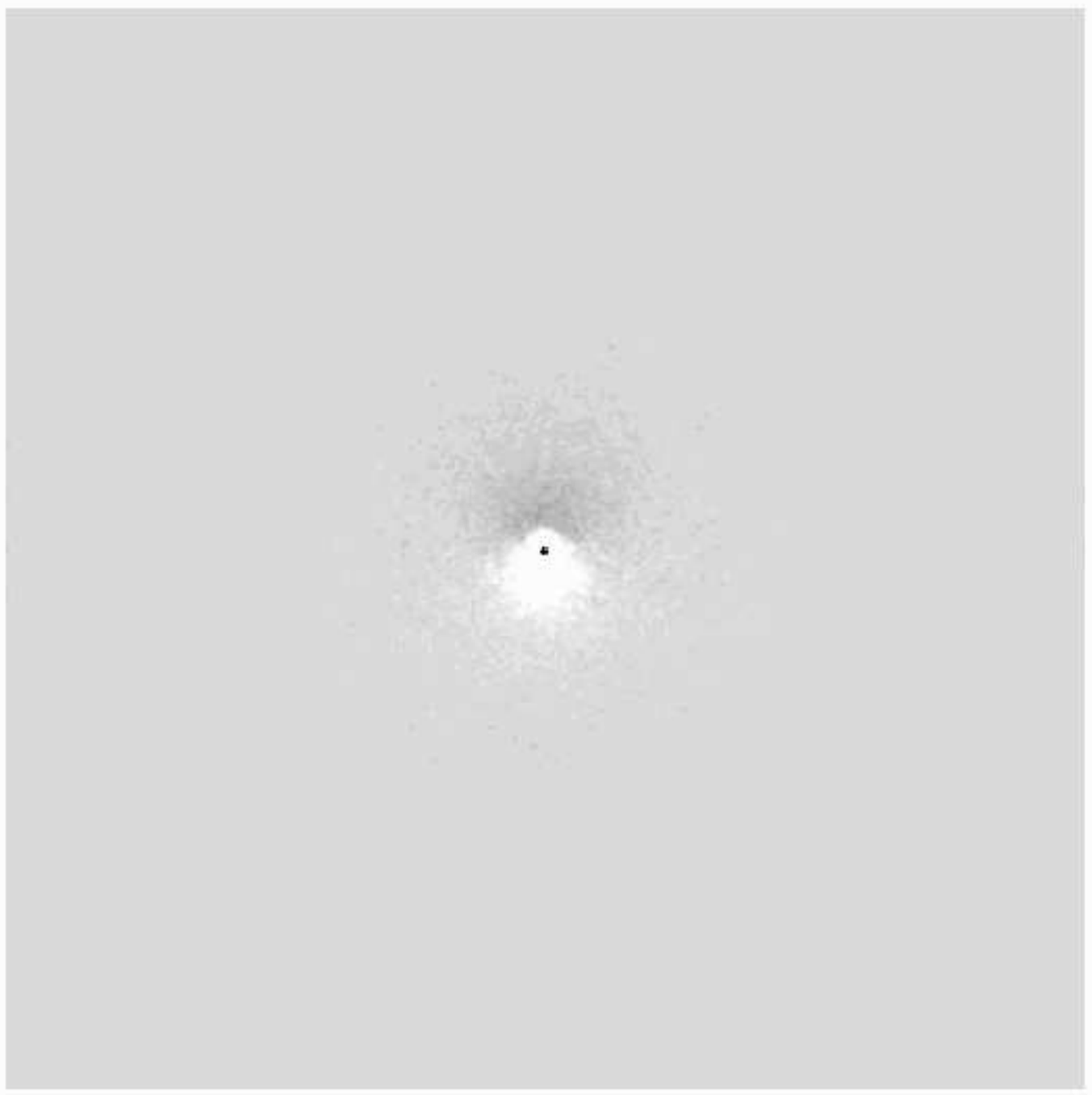}
  \caption{Example of a {\sc budda} fit (middle panel) to a {\sc
      skirt} dusty model (left-hand panel), which corresponds to a
    TrueBulge galaxy model at an inclination of 30 degrees, and with a
    dust opacity of $\tau=1$. The right-hand panel is an enhanced
    residual image, obtained by subtracting the {\sc budda} fit from
    the {\sc skirt} dusty model.  The structure seen in the centre of
    the residual image is mostly a result from attenuation caused by
    the dust disc. The black dot at the centre results from a small discrepancy between the input {\sc skirt} model and the {\sc budda} fit in an area with radius of the order of half the PSF FWHM at the centre of the model images.}
  \label{buddaexample.pdf}
\end{figure*}

We investigate the resulting images from {\sc skirt} using {\sc
  budda}, a code developed to perform a detailed structural analysis
of galaxies through 2D image decomposition, and extensively tested \citep[see][for
details]{2004ApJS..153..411D,2008MNRAS.384..420G}. While the code is
able to fit complex systems such as barred galaxies, here we use it to
fit only bulge and disc, since these are the only components present
in our model images.

The bulge component is fitted with a S\'ersic profile, whereas discs
are infinitely thin exponential discs.  The {\sc budda} code needs a
starting point to fit each parameter, and these values were obtained
using {\sc iraf}. We searched the maximum of the light distribution
and considered this as an input for the centre of the model galaxy.
An intensity radial profile was used to estimate the bulge and disc
structural parameters for each model. The user is also able to set how
much each parameter varies at every iteration in the search for the
best fit. For the disc central intensity and scale length, and for the
bulge effective intensity, effective radius and S\'ersic index, we
typically used a variation of about 10 per cent, as this was shown to
give best results. For the central coordinates and for the geometrical
parameters of bulge and disc (i.e. position angle and ellipticity) we
set this value to $\sim0.1-1$ per cent, as good estimates for these
parameters are relatively easier to obtain, using e.g. the {\sc ellipse}
task in {\sc iraf}. During the fitting process, the code
alters these variations as needed by a factor of 2. For instance, when
approaching the best solution, such variations gradually fall until
convergence is reached. Finally, the fits also take into account the
effects produced by our mimicking of seeing effects (see
Sect. \ref{rts}).  For each model, we have produced and inspected fits
until a satisfactory one was obtained, which usually happened at the first or
second run of {\sc budda}.

Once with the best fit at hand, the next step is to use {\sc budda} to
create an image of the structural model so obtained. To inspect the
quality of the fits we produced residual images by subtracting the
{\sc budda} model image from the {\sc skirt} dusty model image.  An
example of these images is given in Fig. \ref{buddaexample.pdf}.  The
left-hand panel shows an image built with {\sc skirt}, corresponding
to a TrueBulge galaxy model at an inclination of 30 degrees, and with
a dust opacity of $\tau=1$. The middle panel shows the
corresponding {\sc budda} model, whereas the right-hand panel shows
the residual image obtained in this case. While the model images are
displayed using identical brightness and contrast levels, the residual
image is displayed as to significantly enhance the differences between
the two images, which do not go above the 10 per cent level. The structure seen
in the residual image is caused by the dust disc: the {\sc skirt}
model cannot be fitted perfectly anymore by a stellar disc and a
stellar bulge alone.

\section{Results}
\label{sec:res}

In this section, we discuss how the bulge and disc structural
parameters as obtained from the {\sc budda} bulge/disc decompositions
vary with the inclination $i$ and optical depth $\tau$ of
the {\sc skirt} model. The results are summarised graphically in
Figs.~{\ref{DiscParameters.pdf}}
and~{\ref{BulgeParameters.pdf}}. Before discussing the individual
panels in detail, we note that the values of some of the structural
parameters obtained with {\sc budda} at $\tau=0$, i.e. in
the absence of dust, seem to depend on $i$, and not always reproduce
the exact values used as input for {\sc skirt}, although such
differences are very small. The reason for it is that the bulge/disc decomposition
performed by {\sc budda} assumes an infinitely thin disc, whereas the
disc in our input radiative transfer model is a double-exponential
disc with a finite thickness in the vertical direction. With
increasing inclination, the effective thickness of the disc increases
and this influences the results of the bulge/disc decomposition. The
structural parameters affected by this inclination effect are the
bulge effective radius and S\'ersic index, and the disc radial scale length.
Nevertheless, such effect only produces variations which are within the typical
uncertainties estimated by {\sc budda} for each parameter.

As mentioned in the Introduction, we have to distinguish the effects caused by the dust disc on the perceived luminosity and structural parameters of galaxies with single stellar structural components (i.e. either a pure stellar disc or bulge) from the effects on the perceived luminosities and structural parameters of stellar discs and bulges in composite galaxies (i.e. disc galaxies with bulges). One can call the former effects as ``single component effects'', and the latter as ``composite effects''. In some sense, the single component effects are real effects, whereas the composite effects are a combination of these real effects with a bias introduced by the decomposition. Therefore, we will use the following terminology.
We will call the {\em apparent} attenuation of the bulge (or disc) luminosity
that obtained with the {\sc budda} fit, which thus includes not only the
effects of dust on the component luminosity, but also the effects of
dust on the bulge/disc decomposition itself. The apparent attenuation is to be distinguished from what we will call the {\em actual} attenuation. The latter is that in
the {\sc skirt} model, and thus comes from the input model to
{\sc budda}, and refers to single component effects. Note that we will only start discussing the actual attenuation of bulge and disc luminosities in Sect. \ref{sec:discussion}. Thus all instances in which we mention attenuation before Sect. \ref{sec:discussion} refer to composite effects. Furthermore, we only look at single component effects on the attenuation of the integrated bulge and disc luminosities, and not on the other structural parameters, such as the scale lengths and the bulge S\'ersic index. Thus, the dust effects on these structural parameters we study in this paper are always composite effects. 

\subsection{Disc parameters}

\begin{figure*}
  \centering
  \includegraphics[angle=-90,width=0.45\textwidth]{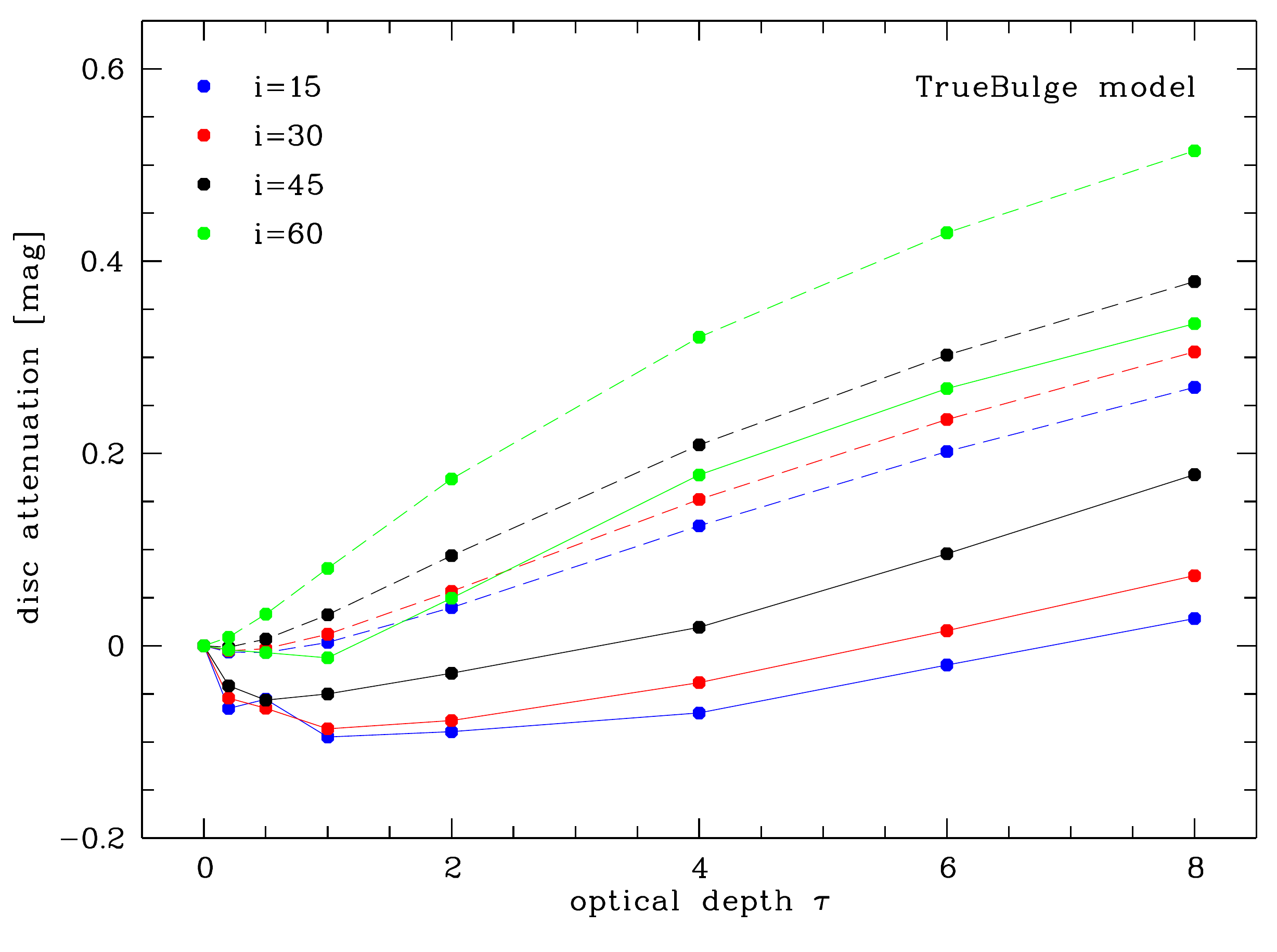}
  \includegraphics[angle=-90,width=0.45\textwidth]{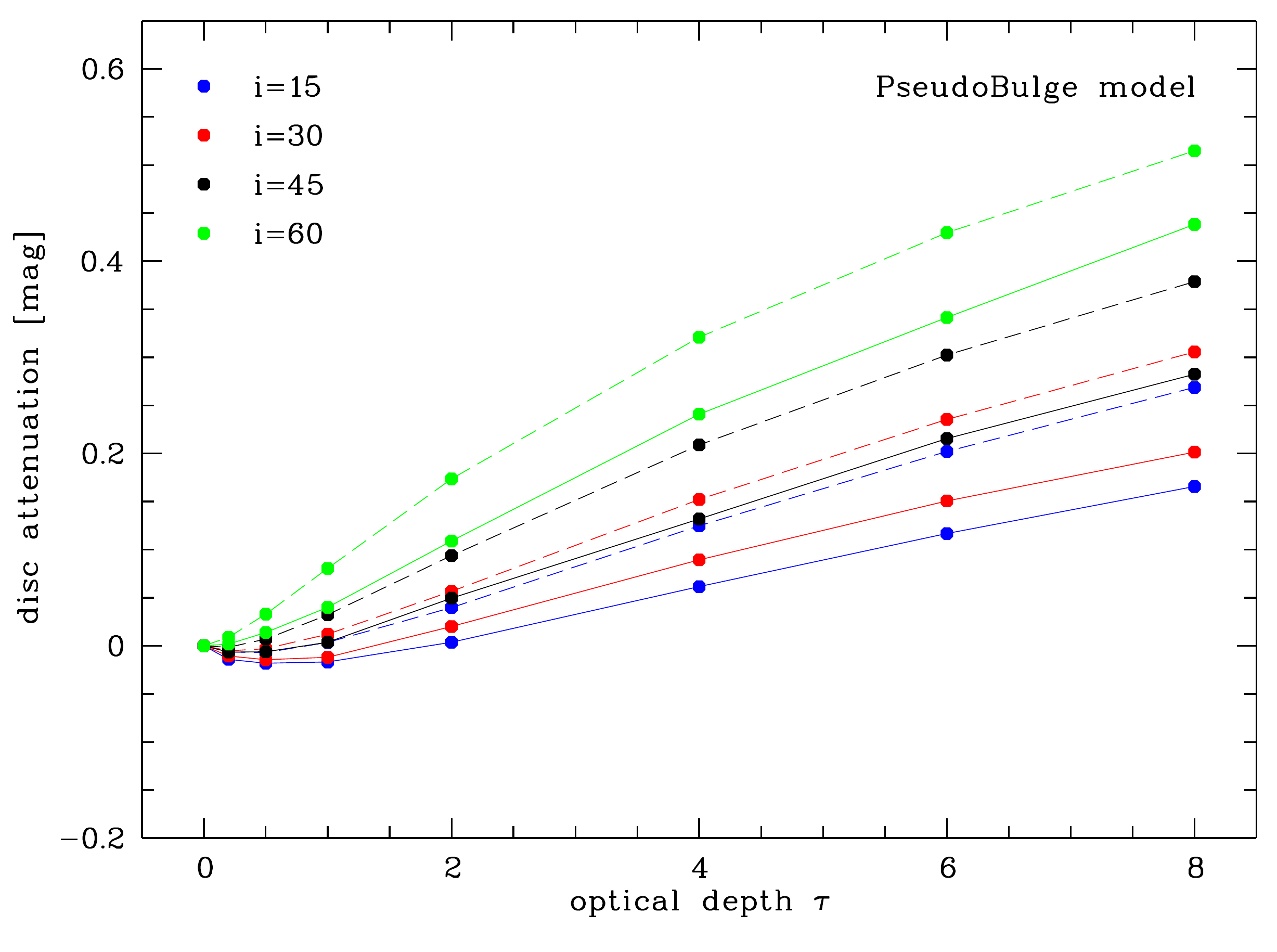}
  \includegraphics[angle=-90,width=0.45\textwidth]{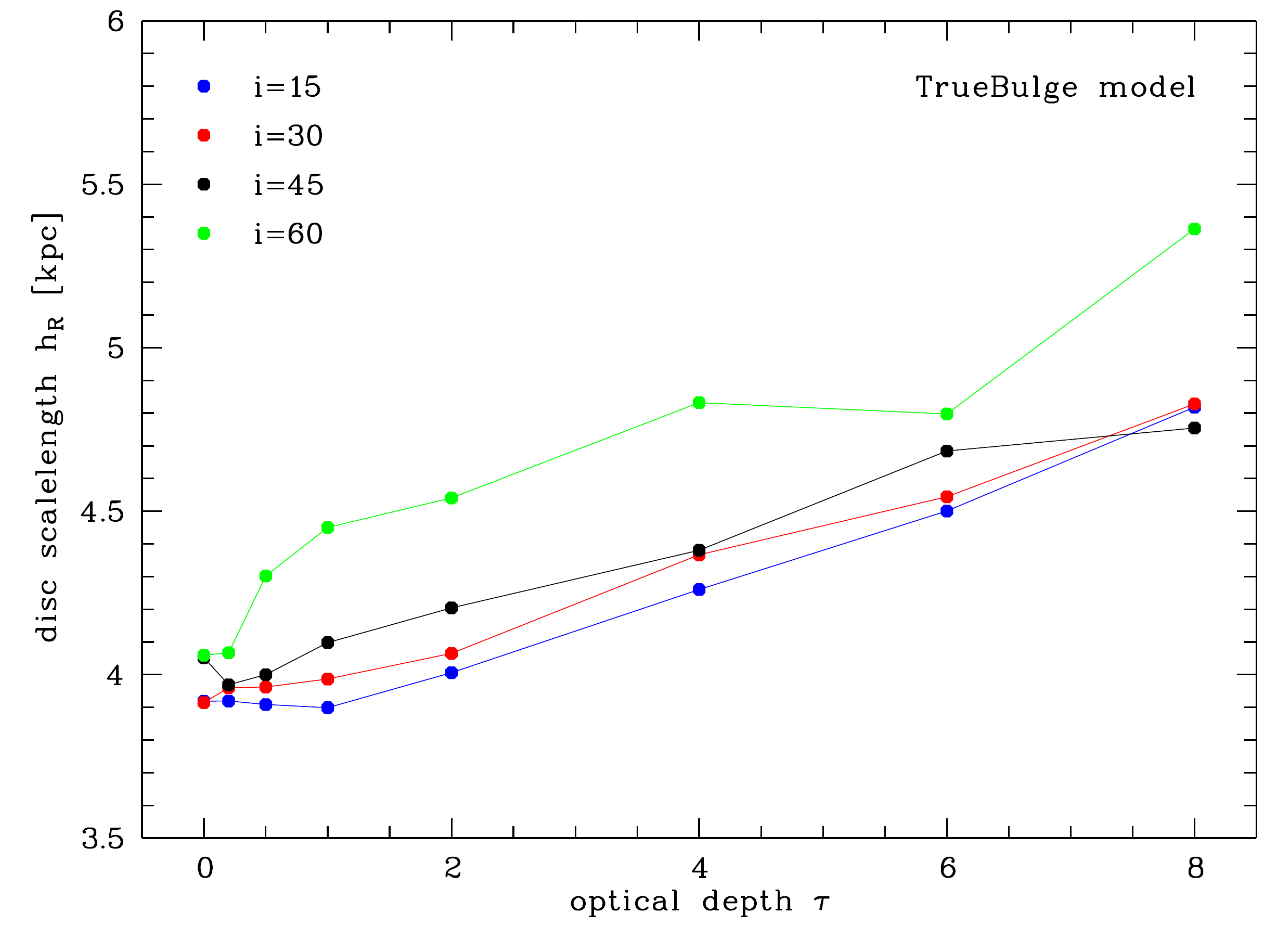}
  \includegraphics[angle=-90,width=0.45\textwidth]{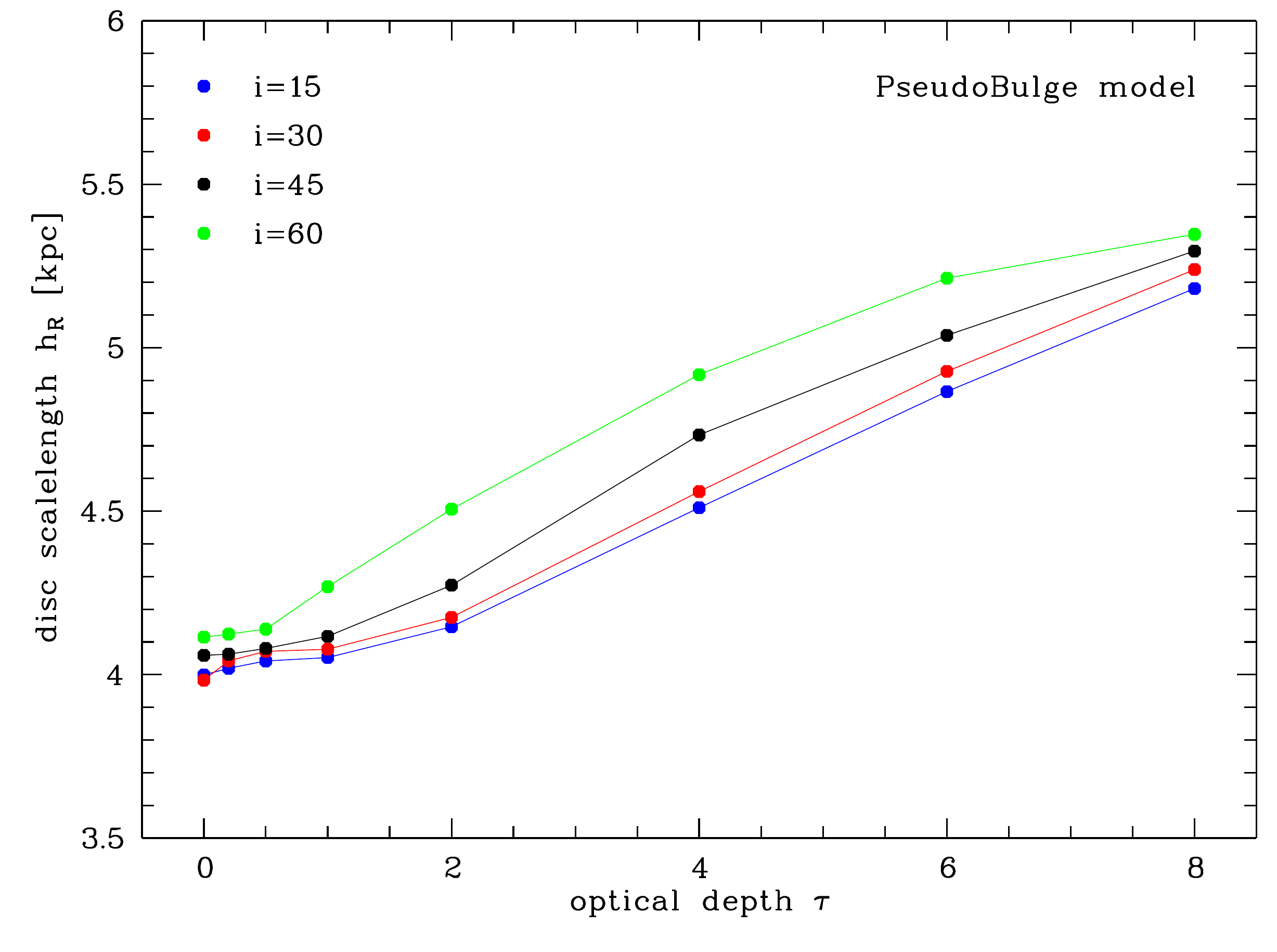}
  \caption{Dependence of the apparent disc parameters on the $V$-band
    optical depth $\tau$, as derived from the {\sc budda}
    bulge/disc decompositions of the dust-affected images. The panels
    on the top and bottom rows respectively show the disc attenuation
    and scale length. The dashed lines in the top panels show the {\em
      actual} attenuation of the disc as a function of the optical
    depth. Results for the different inclination angles are shown, as indicated.}
  \label{DiscParameters.pdf}
\end{figure*}

The solid lines in the top panels of Fig.~{\ref{DiscParameters.pdf}}
show the integrated disc attenuation versus optical depth for the
TrueBulge and PseudoBulge models. Both plots show mostly a
non-monotonous behaviour: the attenuation first decreases to negative
values, reaches a minimum value at $\tau\sim1$, and then
increases to positive values for increasing optical depth. This trend
is clearly present at all inclinations in the TrueBulge models (with a
luminosity increase of up to 0.1 magnitude for the most face-on
inclinations), whereas its effect is much less pronounced in the
PseudoBulge models, being absent in the 60 degrees case.

The positive attenuation of the disc for large optical depths is easy
to understand, although it is not clear at this moment why the attenuation is more
significant in the PseudoBulge models. The negative attenuation, i.e.
the apparent brightening of the disc, at moderate optical depths, on
the other hand, is rather counter-intuitive. One possible explanation
could be the notorious effect of disc brightening at low inclinations
due to dust scattering.  Photons that initially (or after a scattering
event) move on a path nearly parallel to the equatorial plane have a
large probability of interacting with a dust grain. If they are
scattered into a direction that is nearly perpendicular to the disc
plane, the probability of interacting with another dust grain is much
smaller, such that they can easily leave the galaxy.  The overall net
effect of such scattering is that photons are removed from lines of
sight with a high inclination and sent into face-on directions, where
they leave the galaxy. This effect can easily compensate the loss of
radiation due to absorption \citep{1994ApJ...432..114B,
  2001MNRAS.326..733B, 2004ApJ...617.1022P}. However, this effect is
only important for nearly face-on inclinations and cannot play a
significant contribution for apparent disc brightening at inclinations
as high as 60 degrees. Moreover, this effect alone cannot explain the
difference in apparent disc brightening between the TrueBulge and
PseudoBulge models.

The bottom panels of Fig.~{\ref{DiscParameters.pdf}} show the disc
radial scale length $h_R$ versus $\tau$. For all
inclinations and for both the TrueBulge and PseudoBulge models, $h_R$
generally increases with increasing optical depth. This can easily be
understood, given that the dust density decreases exponentially with
increasing radius. The attenuation is therefore stronger in the
central region, which flattens the surface brightness profile and
hence increases the scale length of the best-fitting exponential
profile. However, there is a difference between the TrueBulge and
PseudoBulge models which this effect alone cannot explain: for $i<60$
degrees and $\tau>2$, $h_R$ is systematically larger for
the PseudoBulge models. These results will be further discussed below.

%We have verified that the dust effects on the apparent central surface brightness
%of the disc are essentially similar to those on the integrated disc
%luminosity. They are not exactly the same because disc luminosity also depends
%on $h_R$, but, as we have just seen, dust effects are relatively less dramatic
%for $h_R$.

\subsection{Bulge parameters}

\begin{figure*}
  \centering
  \includegraphics[angle=-90,width=0.45\textwidth]{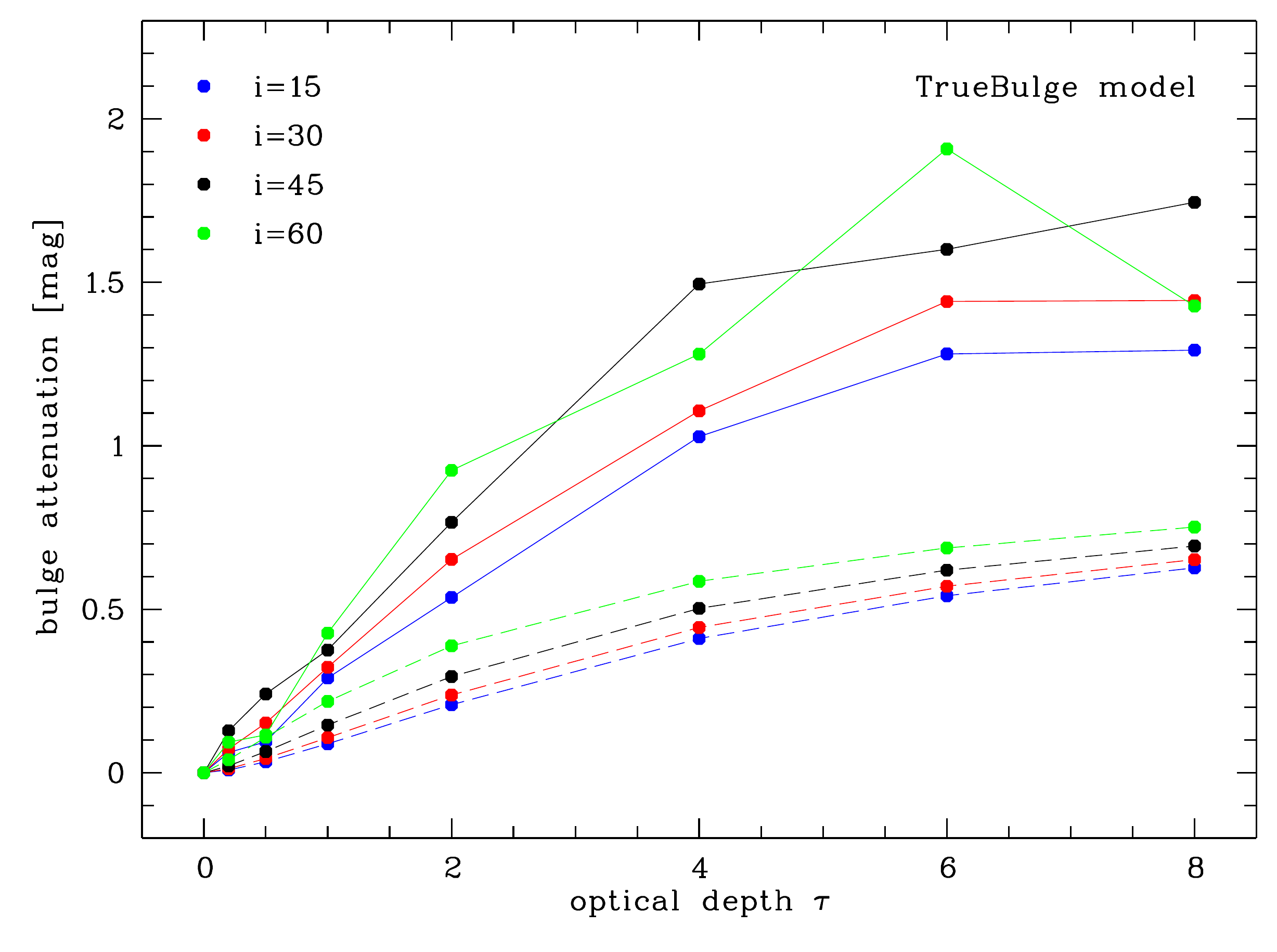}
  \includegraphics[angle=-90,width=0.45\textwidth]{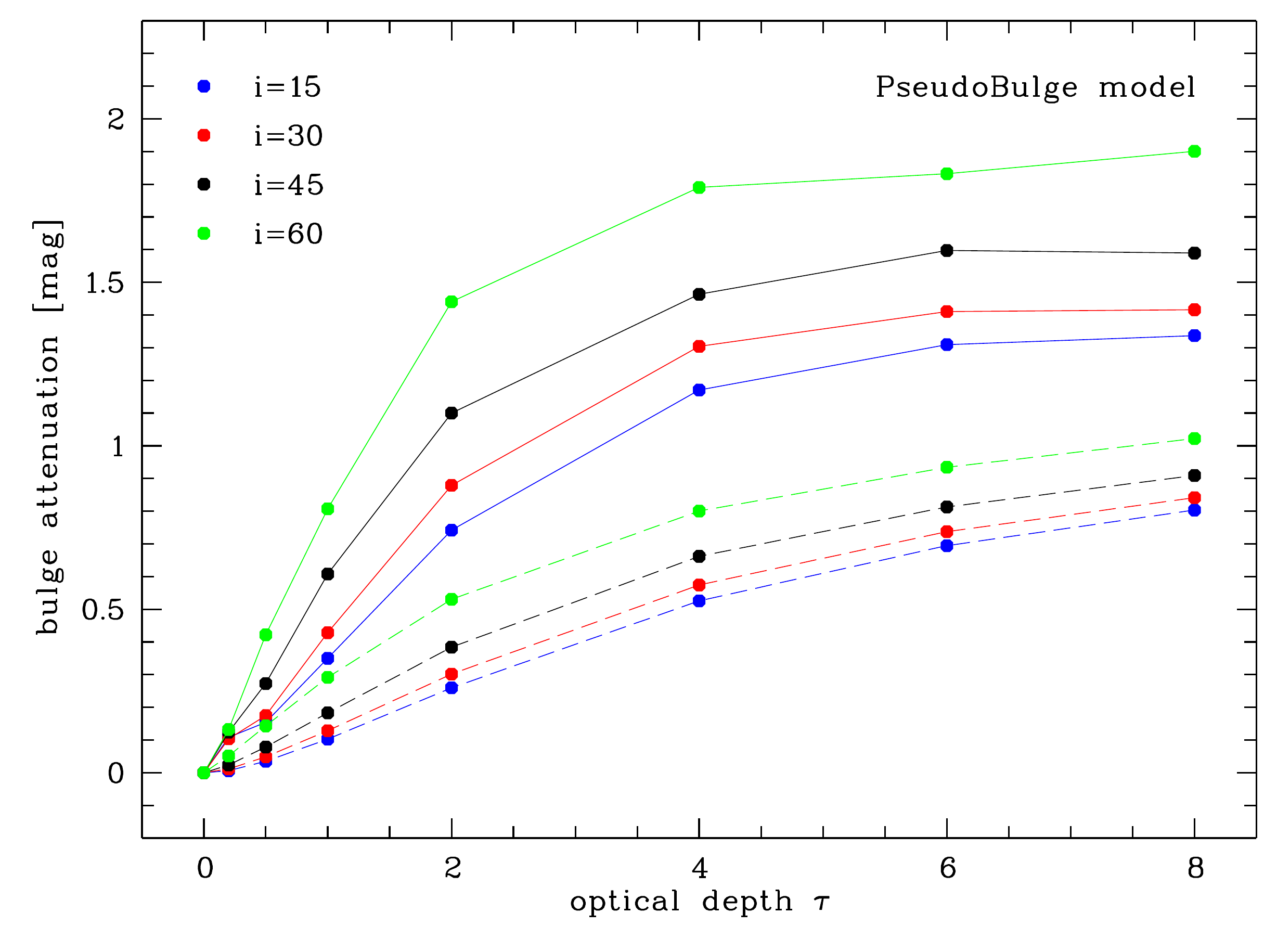}
  \includegraphics[angle=-90,width=0.45\textwidth]{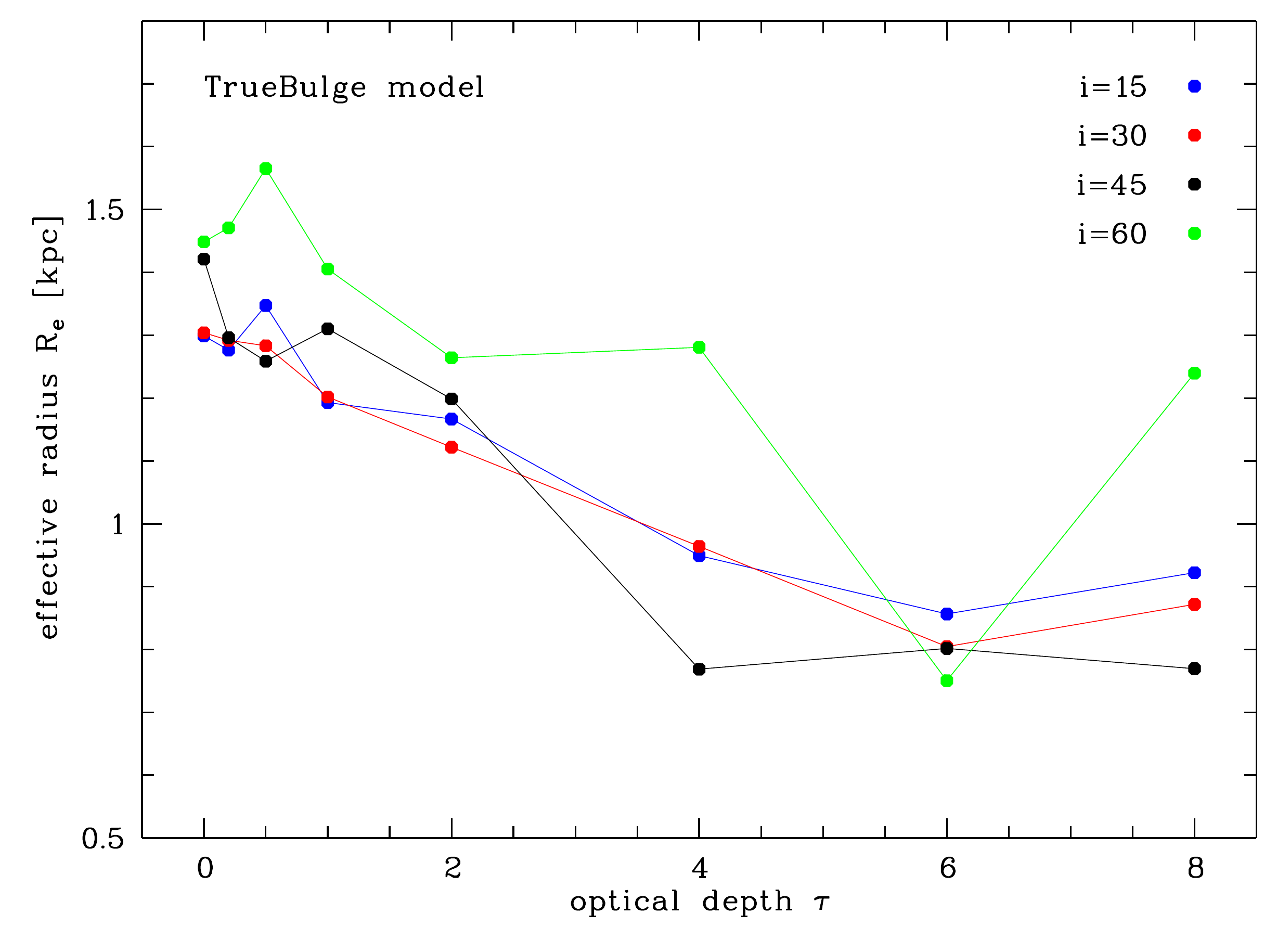}
  \includegraphics[angle=-90,width=0.45\textwidth]{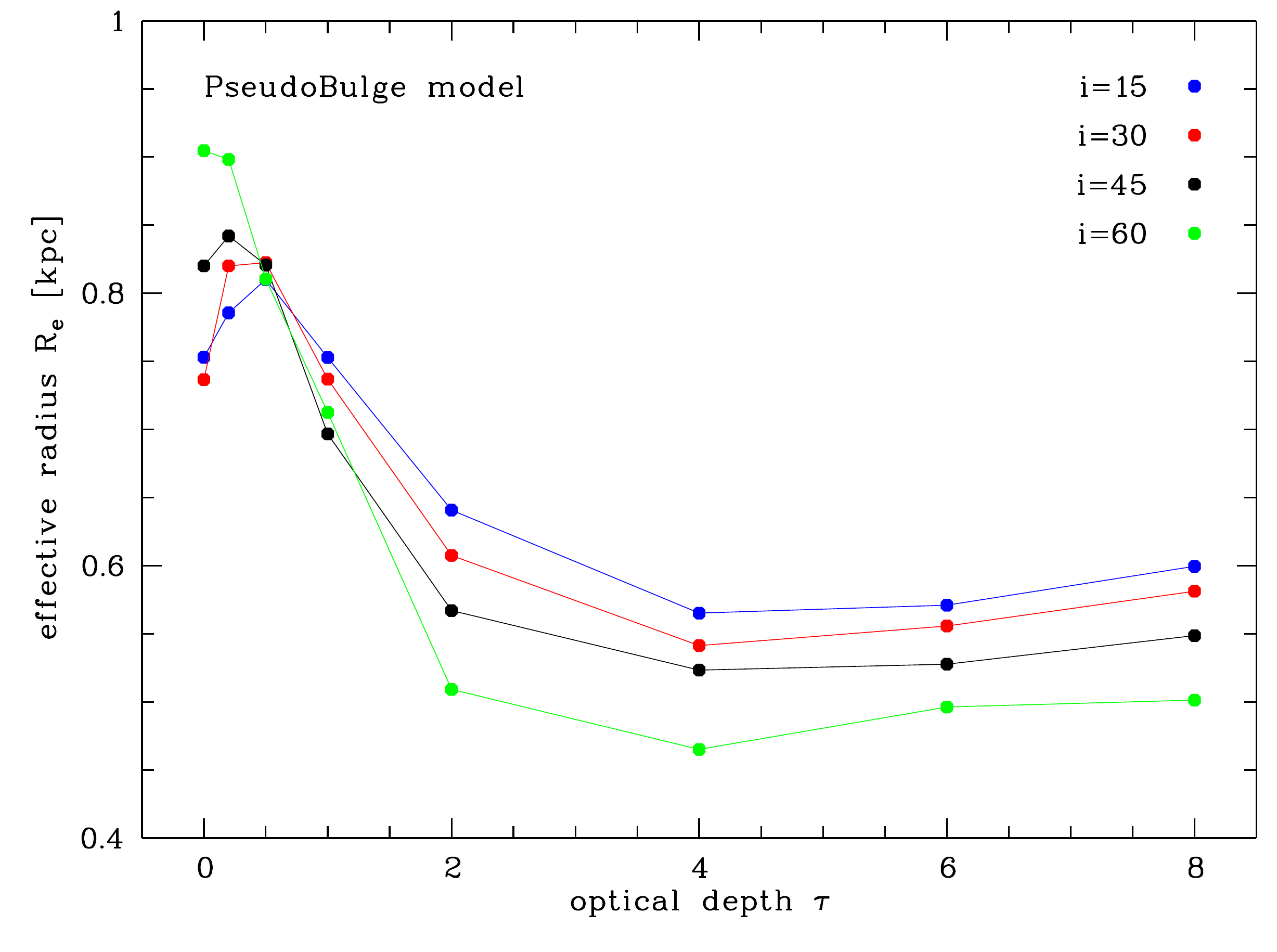}
  \includegraphics[angle=-90,width=0.45\textwidth]{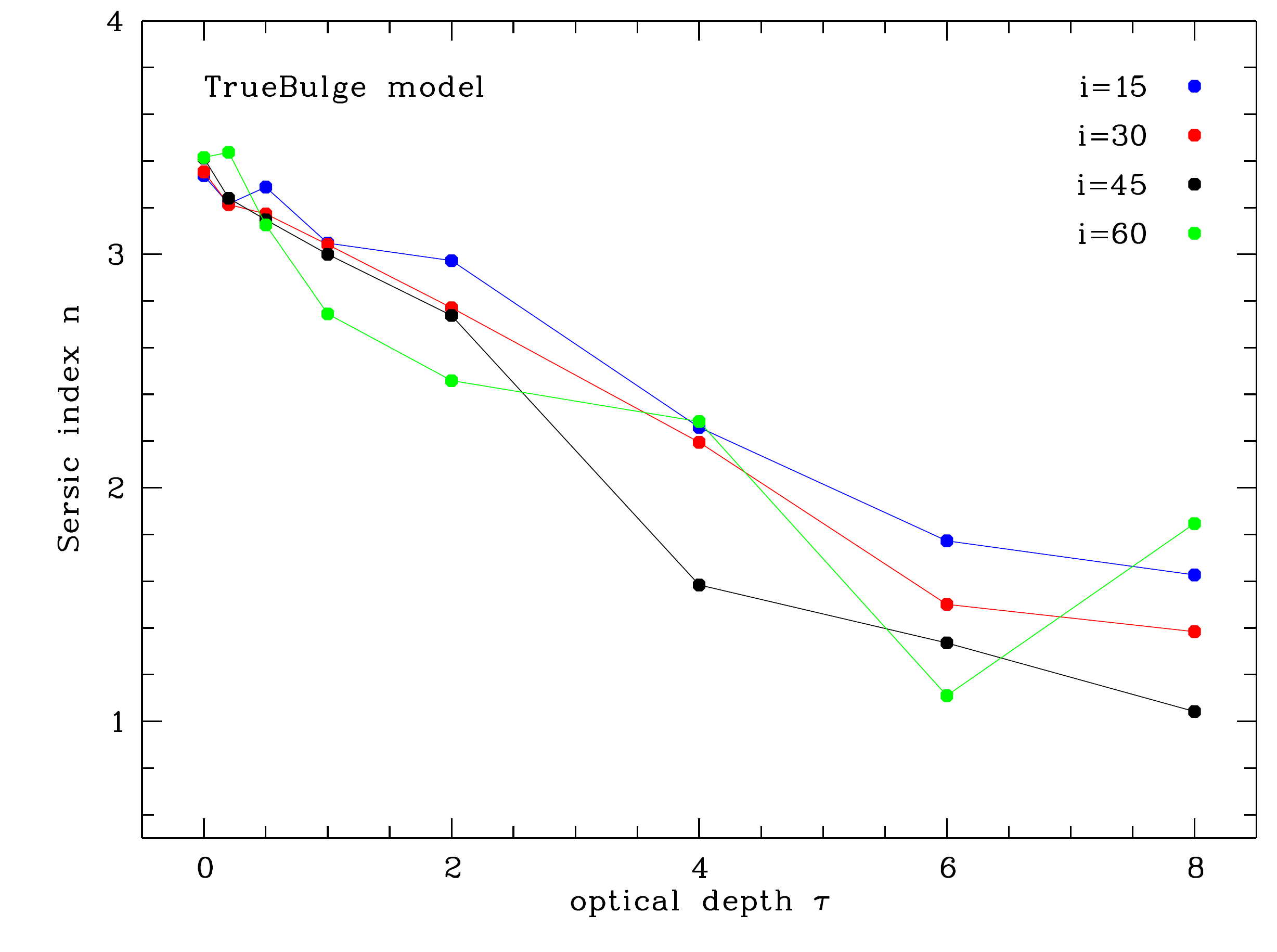}
  \includegraphics[angle=-90,width=0.45\textwidth]{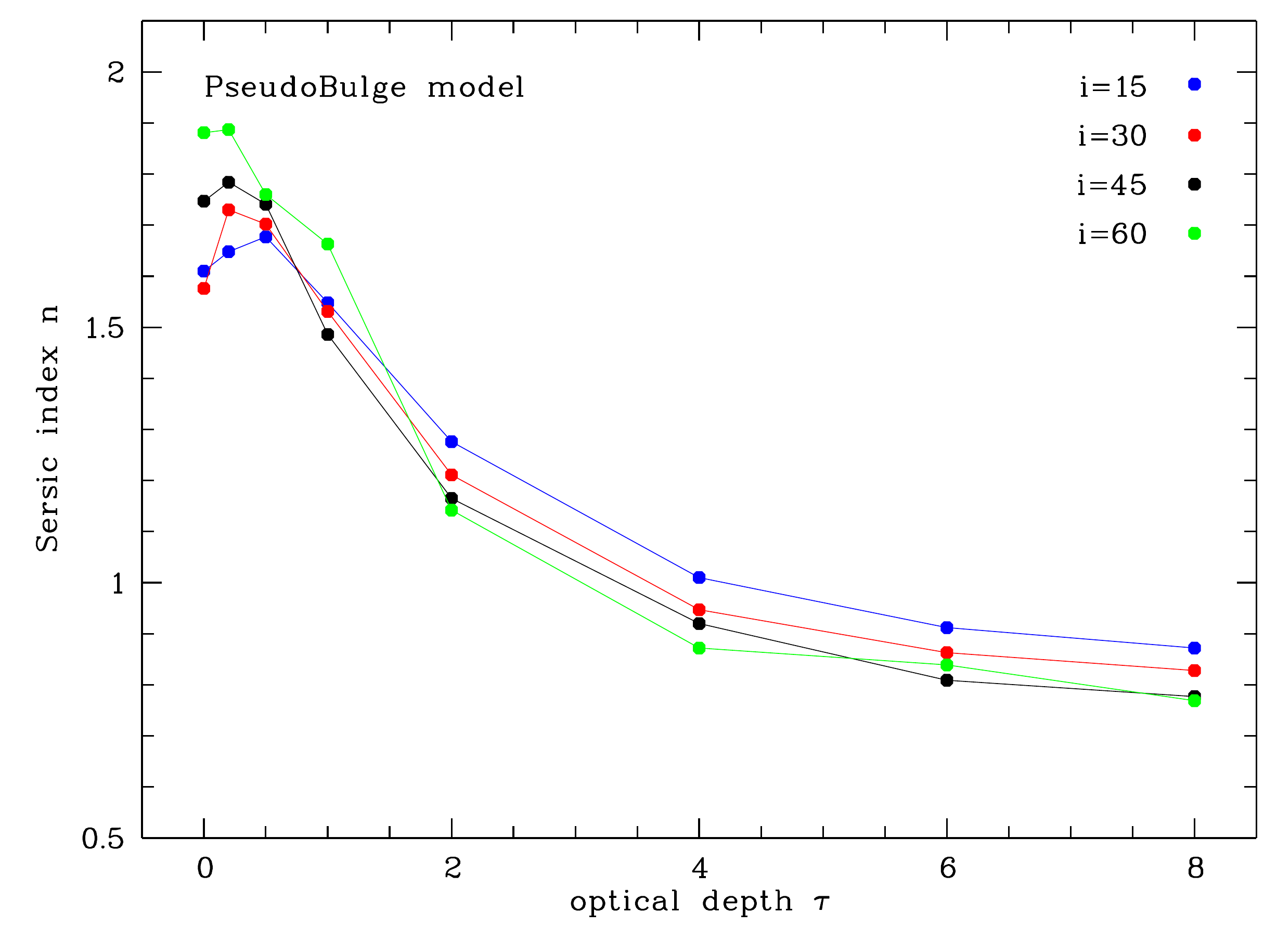}
  \caption{Dependence of the apparent bulge parameters on the $V$-band
    optical depth $\tau$, as derived from the {\sc budda}
    bulge/disc decompositions of the dust-affected images. The panels
    on the top, central and bottom rows show, respectively, the bulge
    attenuation, effective radius and S\'ersic index. The dashed lines
    in the top panels show the {\em actual} attenuation of the bulge
    as a function of the optical depth. Results for the different inclination angles
    are shown, as indicated.}
  \label{BulgeParameters.pdf}
\end{figure*}

The top panels of Fig.~{\ref{BulgeParameters.pdf}} show the integrated
bulge attenuation for the TrueBulge and PseudoBulge models as a
function of optical depth. In both cases, the attenuation increases
drastically with increasing optical depth. It is remarkable that the bulge
attenuation is much stronger than the attenuation of the disc at
identical optical depths and inclinations.

The slopes of the bulge attenuation versus optical depth curves are
not constant, but flatten towards larger optical depths. This is
expected, since at large optical depths the entire central region of
the galaxy (where the bulge resides) becomes completely optically
thick. In these cases, dust behaves as an effective screen for half of
the bulge: the further half of the bulge, i.e. the part that lies
behind the dust layer, does not contribute to the observable bulge
luminosity anymore; only the nearer half of the bulge that lies in
front of the dust layer does contribute. As a result, the attenuation
curves flatten down. This flattening sets in sooner for the
PseudoBulge model (at $\tau\sim4$) than for the TrueBulge
model (at $\tau\sim6$), because the bulge of the former is
smaller in size and hence sooner it reaches the point where the part
of the dust disc covering the bulge is completely optically thick.

The behaviours of the apparent effective radius $R_{\text{e}}$ and
S\'ersic index $n$ are shown in the middle and bottom panels of
Fig.~{\ref{BulgeParameters.pdf}}, respectively.  For the TrueBulge
model, both parameters are generally decreasing functions of the
optical depth. In addition, the corresponding curves also suggest to
be flattening for large values of $\tau$. The more erratic
behaviour of the $i=60$ degrees curve, in this case, is likely due to
the finite vertical thickness of the disc, which is not included in
the {\sc budda} fits. It thus suggests that, for $i\gtrsim 60$
degrees, fits using an infinitely thin disc, as those provided by {\sc
  budda}, start to become unrealistic, since the effective thickness
of the disc is becoming increasingly important.\footnote{Note, however,
  that for edge-on galaxies, i.e. when $i\approx90$ degrees, {\sc
    budda} provides a more realistic fit, using a disc model with finite
vertical thickness.} As a result, the bulge parameters are becoming more
uncertain. For the PseudoBulge model the behaviour is different: both
the effective radius and the S\'ersic index increase with increasing
optical depth, reach a maximum value at $\tau\sim 0.2 -
0.5$ and subsequently decrease for larger optical depths. Only the
effective radius, from $\tau\sim 4$, is roughly constant
with increasing $\tau$, although the curves concerning the
S\'ersic index also flatten for larger optical depths.  It is also
interesting to note how both the apparent effective radius and the
apparent S\'ersic index of the PseudoBulge model coincide at all
inclinations when $\tau=0.5$.

The effect of dust extinction on bulge S\'ersic index can be
intuitively understood.  Since the density in the dust disc is larger
at its centre, the inner part of the bulge surface brightness profile
is the most affected. Therefore, the peak in the centre of the bulge
profile is reduced, leading to lower values of S\'ersic index.  The
flattening of the curves concerning $R_{\text{e}}$ and $n$ also seem
to be related to the same effect leading to the flattening of the
bulge attenuation curves. They set in at similar values of
$\tau$ as the bulge attenuation curves, and also set in for
lower values of $\tau$ in the PseudoBulge models, as
compared to the TrueBulge models.  What is not trivial to understand
is the decrease in effective radius. This leads to a lower
contribution from bulge light also in its outer region, where dust
attenuation is small. Furthermore, it is also unclear why there is an
{\em increase} in both $R_{\text{e}}$ and $n$, in the case of the
PseudoBulge models, when $\tau$ goes from zero to $\sim 0.2
- 0.5$. These points are discussed in more detail in the next section.

%Similarly to what happens with the apparent central surface brightness
%of the disc, we have also verified that the dust effects on the apparent effective surface brightness
%of the bulge are essentially similar to those on the integrated bulge
%luminosity. They are not exactly the same because bulge luminosity also depends
%on $R_{\text{e}}$ and $n$, but, as we have just seen, dust effects are relatively less dramatic
%for $R_{\text{e}}$ and $n$.

\section{Discussion}
\label{sec:discussion}

In the previous section, we have seen that the general behaviour of
the apparent disc and bulge parameters with increasing optical depth
can be understood in an intuitive way, but some results require a more
in-depth discussion:

\begin{itemize}

\item The disc component seems to brighten for moderate optical
  depths, at inclinations which are too far off face-on to explain
  this trend by scattering effects alone. Moreover, the disc
  brightening is stronger in the TrueBulge models than in the
  PseudoBulge models. Scattering effects alone would produce similar
  results, independent on bulge properties.

\item The apparent scale length of the disc is systematically lower
  for the TrueBulge model, as compared with the PseudoBulge model,
  whereas the intrinsic scale lengths are identical.

\item The effective radius of the bulge decreases when dust effects
  become important. However, this leads to a deficit of bulge light in
  its outer region, where dust attenuation is small, as compared to
  the central region.

\item While both the effective radius and S\'ersic index of the bulge
  generally decrease with increasing optical depths, they increase for
  small optical depths for the PseudoBulge models.

\end{itemize}

A plausible explanation for the non-trivial behaviour of these
parameters as a function of optical depth is that they are the result
of the complex interplay between the attenuation of the disc and the
bulge, and its consequences on the 2D bulge/disc decompositions. It is
important to realize that the parameters we study here result from
simultaneous fits to the entire dust-affected image, i.e. to both
bulge and disc at the same time. The apparent bulge parameters are not
only affected by the direct attenuating presence of dust, but also
indirectly by the effect of dust on the apparent disc parameters, and
vice-versa.

In real observations, this complex interplay is difficult to deal
with: photons do not carry a label telling if the star that emitted
them is in the bulge or the disc. In our Monte Carlo models, however,
we {\em can} separate the emission from disc and bulge, and thus study
the effects of dust on each component, individually. In particular, we
can make a subtle but important distinction between the
{\em{apparent}} attenuation of the disc and bulge components and the
{\em{actual}} attenuation, as already discussed at the beginning of Sect. \ref{sec:res}. The apparent attenuation is the measure we
obtain when we consider, as we have done so far, the attenuation
estimated from our bulge/disc decompositions. The apparent attenuation
is thus

\begin{equation}
  A_{{\rm app}} = -2.5\log\left(\frac{L}{L_0}\right),
\end{equation}

\noindent where $L$ is the component integrated luminosity, obtained
from the {\sc budda} decomposition of the dust-affected image, and
$L_0$ is the corresponding luminosity obtained from the decomposition
of the corresponding dust-free image. On the other hand, the {\em
  actual} attenuation is

\begin{equation}
  A_{{\rm act}} = -2.5\log\left(\frac{F}{F_0}\right),
\end{equation}

\noindent where $F$ is the component integrated flux contained in the
dust-affected image and $F_0$ is the corresponding flux in the
optically thin image. In other words, the actual attenuation is that
obtained directly from the {\sc skirt} models, as a result of the
input parameters used to create such models. In the following, we
extend our discussion by considering the actual attenuation in disc
and bulge separately.

\begin{figure*}
  \centering
  \includegraphics[angle=-90,width=0.45\textwidth]{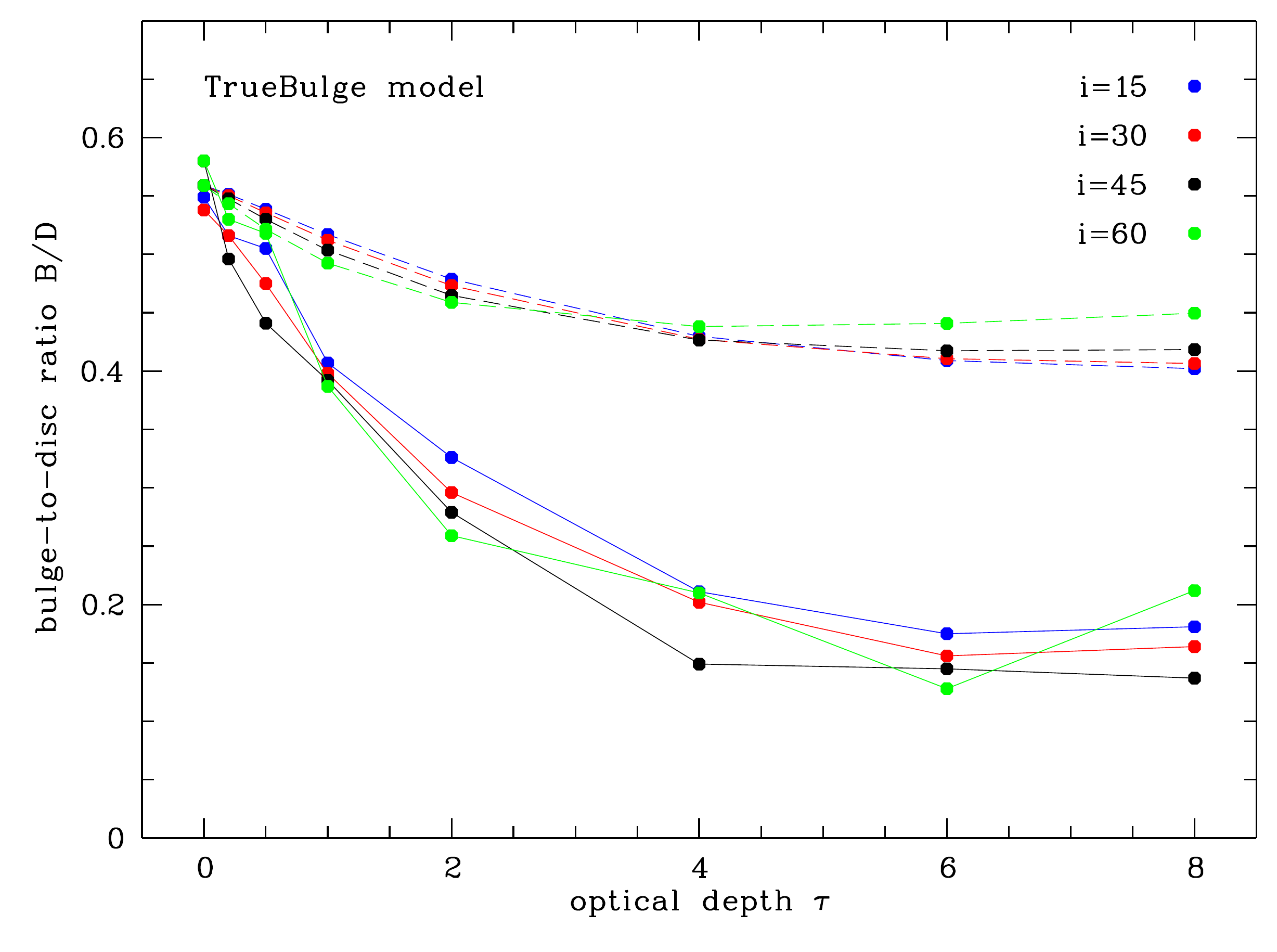}
  \includegraphics[angle=-90,width=0.45\textwidth]{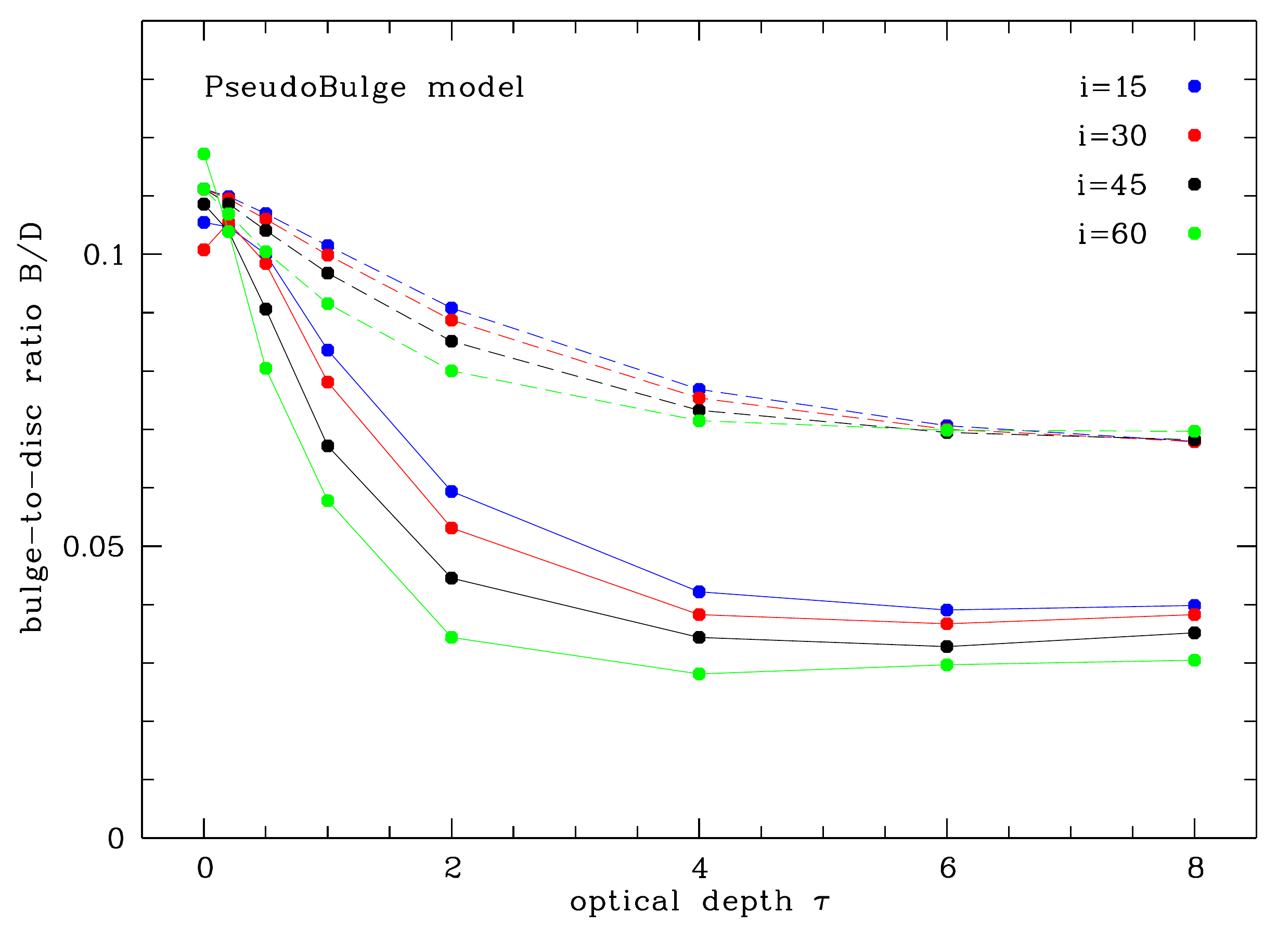}
  \caption{Dependence of the bulge-to-disc ratio on the $V$-band optical
    depth $\tau$. The solid lines represent the apparent
    bulge-to-disc ratio as derived from the {\sc budda} bulge/disc
    decompositions of the dust-affected images. The dashed lines
    represent the actual bulge-to-disc ratio as determined from the
    ratio of the input bulge and disc integrated fluxes.}
  \label{BulgeToDiscRatio.pdf}
\end{figure*}

\subsection{Disc attenuation}

The dashed lines in the upper panels of
Fig.~{\ref{DiscParameters.pdf}} represent the actual disc attenuation
as a function of $\tau$, whereas the solid lines represent
the apparent attenuation. We note in both cases the intriguing disc
brightening, but this effect is only found for the more face-on
inclinations when we consider the actual attenuation; further, the
effect is identical in the TrueBulge and PseudoBulge models. It can be
completely understood in terms of scattering effects
\citep{1994ApJ...432..114B, 2001MNRAS.326..733B,
  2004ApJ...617.1022P}. This confirms our previous argumentation that
the {\em apparent} disc brightening (the solid lines) cannot be due to
scattering effects alone.

Another difference between the solid and dashed lines is the strength
of the attenuation: the apparent disc attenuation is significantly
weaker than the actual attenuation at a given $\tau$. In
other words, the bulge/disc decomposition minimizes the attenuation of
the disc. The reason for this difference stems from the different ways
the inner and outer regions of the disc relate to the resulting
apparent and actual attenuation. The outer regions give only a modest
contribution to the total flux of the disc; the inner, dust-affected
regions contribute much more significantly, and are mainly responsible
for the general increase of the actual attenuation with increasing
optical depth. On the other hand, it is the outer regions of the disc
that have a relatively larger weight on the resulting apparent disc
attenuation.  In fact, when searching for the best disc model, {\sc
  budda} naturally gives more weight to the outer disc, as compared to
the inner disc, as the former occupies a much large number of pixels
in the image.\footnote{The weighting scheme used in {\sc budda} is such that each pixel is weighted by a function which is $\propto1/N$, where $N$ in the number of counts in ADU at the pixel, i.e. inversely proportional to the absolute value of the Poisson noise (variance) in the pixel. This too tends to put more weight on the outer parts of galaxies, since pixels there have generally lower counts.}  Furthermore, the outer surface brightness profile of
the disc is hardly affected by dust.  As a result, the outer regions
will tend to push the fitted disc parameters to their optically thin
value, in spite of the much larger attenuation at smaller radii,
making the apparent attenuation weaker than the actual one. As we will
discuss now, in order to compensate such discrepancy at small radii,
the bulge model is affected in such a way that the bulge contribution
to the total model is reduced.

\subsection{Bulge attenuation}

\begin{figure}
  \centering
  \includegraphics[angle=0,width=0.45\textwidth,clip=true]{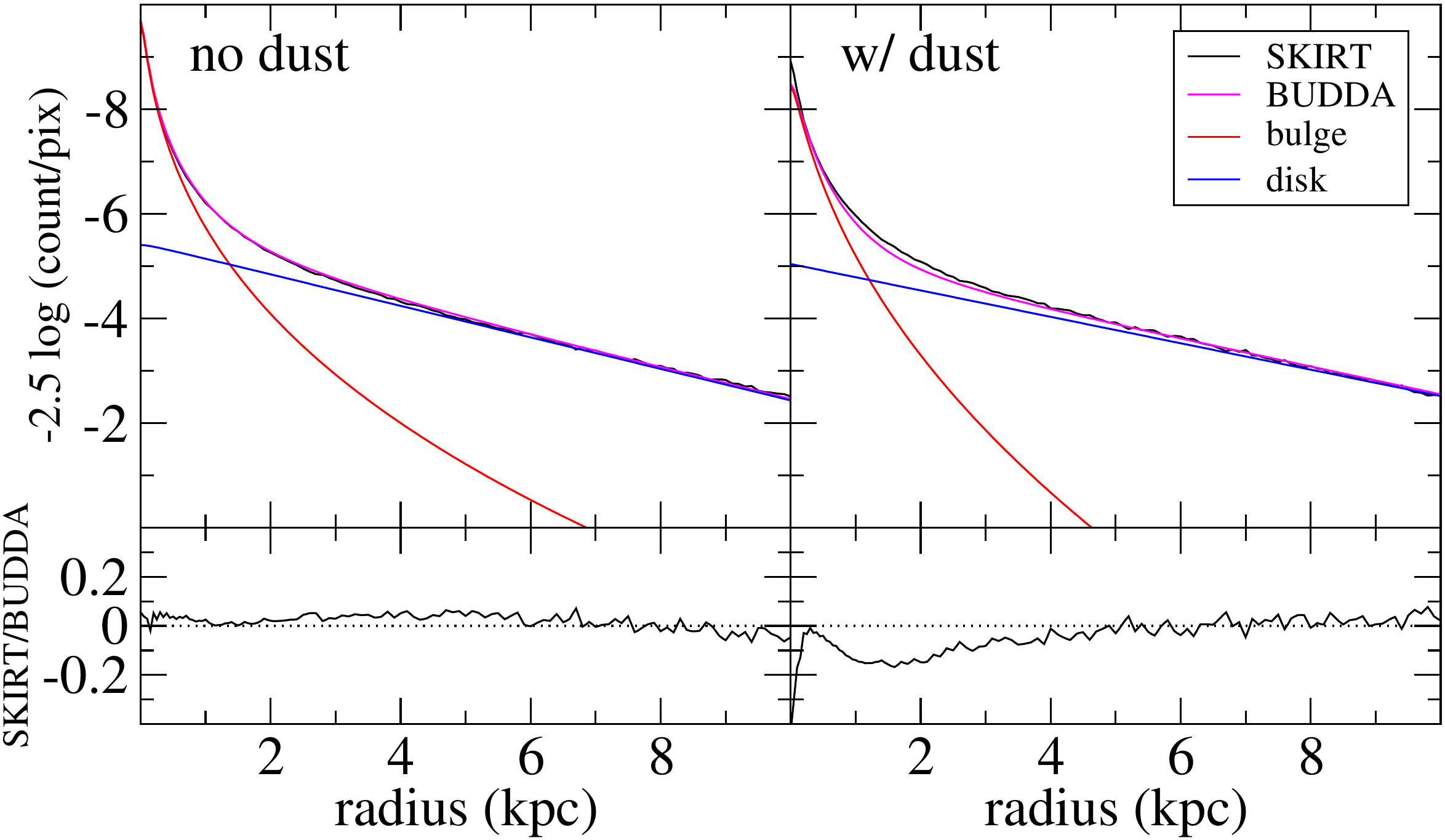}
  \caption{Results from {\sc budda} decompositions of two TrueBulge {\sc skirt} models seen at a viewing angle $i=60^\circ$, with no dust (left) and with a dust disc with $\tau=4$ (right). The surface brightness radial profiles of the {\sc skirt} and total {\sc budda} models are shown in the upper panels, as well as those of the bulge and disc individual {\sc budda} models, as indicated. The lower panels show the residuals of the fits in magnitudes. One can clearly see how dust effects are more severe in the bulge model than in the disc model.}
  \label{referee}
\end{figure}

A direct consequence from the above is thus that the bulge parameters will be
strongly affected. This is clear when we compare the solid and dashed
lines in the top panels of Fig.~{\ref{BulgeParameters.pdf}}, which
represent the apparent and actual bulge attenuations,
respectively. Clearly, the apparent bulge attenuation at a given
optical depth is much stronger than the actual bulge attenuation,
contrary to what happens with the disc apparent and actual
attenuations. The magnitude of this difference is dependent on the
inclination of the system and the type of bulge, but it typically
amounts to a factor of two or more. This is completely in accordance
with the previous observation that, in the bulge/disc decomposition,
to account for the excess of light from the disc model in the galaxy
inner region, the bulge model becomes less important. This explains
why the effective radius of the bulge is also reduced as dust effects
become important. Further, as a consequence, the bulge-to-disc
luminosity ratio, as measured through the decompositions, can be
significantly smaller than the corresponding ratio in the input, {\sc
  skirt} models, at fixed optical depth and inclination (see
Fig.~{\ref{BulgeToDiscRatio.pdf}}).

This effect arising from the bulge/disc decompositions also explains
why the disc suffers more from dust effects in the PseudoBulge models,
as compared to the TrueBulge models.
Figure~{\ref{DiscParameters.pdf}} shows that the effects of dust on
the apparent integrated disc attenuation and scale length are more
significant in the PseudoBulge models. What happens is that because
the bulge in the PseudoBulge models is relatively small, it cannot be
weakened so significantly, as the bulge in the TrueBulge models, to
account for the reduced effects of dust in the disc. Therefore, the disc itself,
in the PseudoBulge models, has to respond more substantially
to the dust effects.

To better illustrate the effects of dust in the {\sc budda} fits, we show in Fig.~{\ref{referee}} the results from {\sc budda} decompositions of two TrueBulge {\sc skirt} models seen at a viewing angle $i=60^\circ$, with no dust (left-hand panels) and with a dust disc with $\tau=4$ (right-hand panels). While the effects of dust in the disc model obtained with {\sc budda} are relatively weak -- the disc central surface brightness gets slightly fainter and the disc scale length slightly larger -- such effects are strong in the corresponding bulge model, which has both its effective radius and S\'ersic index significantly lowered, resulting in an underestimation of the true bulge-to-disc ratio.
Furthermore, it is also noticeable that the {\sc budda} fit is worse when dust effects are important, in particular at the central regions. In extreme cases, this could lead to the wrong conclusion that substructures are present.

It is not clear, however, how the disc brightening seen above can
happen at high inclinations, and why it is stronger in the TrueBulge
models. We also do not have a clear answer to why, in the PseudoBulge
models, the bulge effective radius and S\'ersic index increase at
small optical depths. Nevertheless, we suspect that both features are
connected, and have an origin in the process of bulge/disc
decomposition.

\section{Implications}

Our results indicate that, broadly speaking, when dust effects are
important, disc scale lengths are being overestimated, and bulge
effective radii and S\'ersic indices and bulge-to-disc ratios
underestimated, in studies of galaxy structure based on bulge/disc
decomposition. The extent of the problem of course depends on galaxy
inclination and dust opacity, and the latter is most of the times
difficult to estimate. Furthermore, while dust effects might not be
too significant on disc parameters, the impact on bulge parameters can
be considerable.  Finally, it is also worth mentioning that, for high
opacities, dust effects might not be negligible even for galaxies seen
face-on.

It is well known that dust attenuation in galaxies also depends strongly on
the dust-star geometry \citep{1989MNRAS.239..939D,
  1992ApJ...393..611W, 1994ApJ...432..114B, 1996ApJ...465..127B,
  2001MNRAS.326..733B}. Applied to the particular case of spiral
galaxies containing both a disc and a bulge component, it is not
surprising that interstellar dust affects the bulge much stronger than
the disc. The fact that bulges suffer more attenuation than discs has
been demonstrated in detail by several authors
\citep{2004ApJ...617.1022P, 2004A&A...419..821T,
  2006A&A...456..941M}. Using radiative transfer calculations that
investigate the effects of dust on the actual individual bulge and disc
components, \citet{2004A&A...419..821T} concluded that increasing the
bulge-to-disc ratio at a constant opacity can mimic the effect of
increasing the opacity of a pure disc. In general, ignoring the
presence of bulges can lead to a systematic overestimate of the
opacity of discs.

In the previous section, we have demonstrated that the actual and
apparent attenuations of disc and bulge are indeed substantially
different: the actual disc attenuation is modest (even negative for
small optical depths due to scattering effects), whereas the actual
bulge attenuation is much more pronounced. This differential
extinction between bulge and disc leads to a general decrease of the
bulge-to-disc ratio with increasing optical depth, as demonstrated
with the dashed lines in Fig.~{\ref{BulgeToDiscRatio.pdf}}. For the
TrueBulge model, the actual bulge-to-disc ratio decreases some 10 to
15\% at an optical depth $\tau=2$, reaching a reduction of
roughly 25\% at $\tau=8$. For the PseudoBulge model, with
its smaller bulge buried deeper into the dusty central region, the
actual bulge-to-disc ratio decrease is slightly stronger, being about
20\% at $\tau=2$ and 35\% at $\tau=8$.

While these numbers are significant, the actual extinction of bulge
and disc light is not the dominant factor in the behaviour of the
apparent bulge-to-disc ratio with increasing dust opacity. The solid
lines in Fig.~{\ref{BulgeToDiscRatio.pdf}} represent the apparent
bulge-to-disc ratio resulting from the 2D bulge/disc decomposition of
the dust-affected images. It is important to stress that this measure,
and not the actual bulge-to-disc ratio discussed previously, is the
observable bulge-to-disc ratio measured in real galaxy images. These
curves are the result of two corroborating effects: on the one hand
the different actual attenuation of the individual bulge and disc
components, and on the other hand the effect with origin in the
process of bulge/disc decomposition, in which the bulge model is
reduced in order to compensate for an excess of light in the central
part of the disc model, which in the actual disc is dust-obscured.  We
have seen in the previous section that this effect tends to minimize
the attenuation of the disc while it strongly overestimates the
attenuation of the bulge. The result is that the apparent
bulge-to-disc ratio is a stronger function of the optical depth than the
actual bulge-to-disc ratio:
even at relatively modest optical depths, $\tau=1$ and $\tau=2$, the apparent
bulge-to-disc ratio decreases by roughly 25\% and 50\%
(or bulge-to-{\em total} ratio decrements of roughly 15\% and 33\%), respectively,
for {\em both} the TrueBulge and PseudoBulge cases. Interestingly enough,
also for both types of models, the apparent bulge-to-disc ratio
becomes roughly constant for values of $\tau$ larger than 4. At this
point, the underestimation of the bulge-to-disc ratio is about a factor
of 3, which corresponds to an underestimation of the bulge-to-total ratio
by a factor of 2. Such effect can have important consequences on studies
of the black hole/bulge scaling relations in disc galaxies. Some such studies
\citep[see e.g.][and references therein]{2009ApJ...698..198G,2009ApJ...694L.166B} derive
bulge luminosity or mass from the bulge-to-total ratio, which, as we have seen,
can be substantially underestimated if dust attenuation is significant. This is likely to happen
if the image used in the decomposition is obtained through a blue passband, or if
the galaxy is seen edge-on. An artificial decrement in bulge-to-total ratio can result
in an elevated measure of the scatter in such relations, and one thus has to be alert
to that possibility.

Given the difficulties in determining dust opacities, we will not attempt to draw
quantitative conclusions about the effects of dust on the outcome of observational
work on the structural parameters of galaxy components through image
decomposition. However, we note that, knowing the galaxy inclination,
and assuming or estimating a value for $\tau$, one
can roughly assess in the figures presented above how the galaxy
structural parameters are affected through 2D bulge/disc decomposition.

The biases produced by dust effects on the results from decompositions in studies with statistically large samples will depend on the spatial resolution of each object, as well as on details in the decomposition algorithm. Our results indicate that bulges which are better resolved will tend to suffer more from dust effects. In addition, the function used to describe the surface brightness radial profile of discs, and the weighting scheme used by the decomposition algorithm, also influence how dust affects the results. Using a broken exponential profile to fit discs, with the inner part flatter than the outer part, might alleviate biases caused by dust. This can also be achieved by given more weight to pixels in the inner part of the disc, as compared to its outer part.\footnote{Note, however, that different weighting schemes can introduce other sorts of biases \citep[see discussion in][]{2005MNRAS.362.1319L}.} One can devise more elaborated fitting functions or weighting schemes if the disc opacity can be evaluated by independent means, through e.g. measurements of dust emission and/or spectroscopic diagnostics. Thus, biases caused by dust have to be calibrated separately for each survey.

Since bulge prominence is an important structural property, connected
to other important physical properties, one should be careful when
assessing bulge-to-disc ratios (or bulge-to-total ratios) in cases where dust effects are
expected to be significant. The results above show that the effects of dust
on the measured structural parameters of bulges and discs can be different in galaxies
hosting classical and pseudo-bulges, even if their dust content is the same.
It might be thus fortunate that, when such effects are combined, they result in
relative decrements in the bulge prominence which are quantitatively similar for both
categories of galaxies.

\section{Summary and conclusions}
\label{concl.sec}

We have created artificial galaxy images, using radiative transfer
simulations, to mimic the observed structural properties of galaxies
with classical and pseudo-bulges, and include the effects of dust
attenuation in the observed light distribution. By applying 2D
bulge/disc decomposition techniques in this set of models, we were
able to evaluate what are the effects of galaxy inclination and dust
opacity on the results from such decompositions.

We have found that the effects of dust on the structural parameters of
bulges and discs obtained from 2D bulge/disc decomposition cannot be
simply evaluated by putting together the effects of dust on properties
of bulges and discs treated separately. The reason for that comes from
the fact that such decompositions use specific models to fit bulges
and discs which cannot accommodate the effects of a dust disc in the
galaxy. Therefore, when the model for a component tries to adjust
itself when dust is present, this has direct consequences on the model
of the other component, even if the latter is not directly affected by
dust. More specifically, we have found that the outer parts of the
disc have more influence in setting the disc model in the 2D fits, as
they occupy a larger number of pixels. However, this area of the disc
suffers little from dust attenuation, and hence the disc parameters so
obtained are not significantly affected by dust. This results from the
fact that a simple exponential function is used to describe the light
distribution in the disc. As a consequence, there will be an excess of
light in the central parts of the disc, because dust attenuation is
stronger there. Finally, to compensate such discrepancy in the central
parts of the galaxy, the bulge model is reduced. A possible solution
to such problem would thus be to use a broken exponential to fit the
disc light distribution, with the inner part of such function being
flatter than the outer part. Another possible solution is to use a pixel weighting scheme in the decomposition algorithm that compensates for the effects of dust.

We have found that, when dust effects are important, disc scale
lengths are overestimated, bulge effective radii are underestimated,
and bulge S\'ersic indices are also underestimated. Furthermore, the
attenuation caused by dust in the integrated disc luminosity is
underestimated, whereas the corresponding attenuation for the bulge is
overestimated. This leads to a systematic underestimation of
bulge-to-disc ratios.  The extent to which these systematic effects
are significant depend on galaxy inclination and dust opacity. The
strongest effect is seen in the bulge-to-disc ratio, which can be
underestimated by a factor of two, in the $V$ band, even considering relatively low
inclinations and opacities. Nevertheless, we have also found that such
parameter is never underestimated by factors larger than about three,
which corresponds to a factor of two in bulge-to-total ratio. Such effect
can have an impact on studies of the black hole/bulge scaling relations.

\section*{Acknowledgments}
We are grateful to Daniele Pierini for reading our manuscript and providing
insightful comments. We thank an anonymous referee for several comments that helped to improve our paper. DAG is supported by the Deutsche Forschungsgemeinschaft priority
program 1177 (``Witnesses of Cosmic History: Formation and evolution
of galaxies, black holes and their environment''), and the Max Planck
Society. MB and SF gratefully acknowledge the financial support of the
Fonds voor Wetenschappelijk Onderzoek Vlaanderen (FWO-Vlaanderen).

\appendix
\section{S\'ersic models in SKIRT}

The surface brightness profiles of bulges in spiral galaxies can
generally be represented as S\'ersic models, a convention that is
followed in this paper. One problem with the S\'ersic models is that
the deprojection of a S\'ersic surface brightness profile to a volume
emissivity $\nu({\bf{r}})$ is in general non-analytic and that it is
singular at the centre. In practice, one often uses approximations for
the S\'ersic models when the volume emissivity is necessary
\citep[e.g.][]{1997A&A...321..111P, 1999MNRAS.309..481L,
  2002MNRAS.333..510T}. Since we model the projected images of our
{\sc{skirt}} models with exact S\'ersic models, it is important that
the correct, i.e.\ exact, volume emissivity of the S\'ersic models are
adopted in this case.

Fortunately, divergent volume emissivity profiles do not pose a
problem for {\sc{skirt}}, as long as the total luminosity is
finite. When running a {\sc{skirt}} Monte Carlo radiative transfer
simulation, the volume emissivity $\nu({\bf{r}})$ does not need to be
calculated explicitly; one only needs to generate random positions
${\bf{r}}$ from the appropriate distribution function,
\begin{equation}
  p({\bf{r}})\,{\text{d}}{\bf{r}}
  = 
  \frac{\nu({\bf{r}})\, {\text{d}}{\bf{r}}}{L},
\end{equation} 
where $L = \int \nu({\bf{r}})\, {\text{d}}{\bf{r}}$ is the total luminosity of the
system. For a spherically symmetric model, such as we have assumed
here, generating a random position ${\bf{r}}=(r,\theta,\phi)$ comes
down to generating a random direction $(\theta,\phi)$ on the unit
sphere and a random radius $r$ from the probability distribution
\begin{equation}
  p(r)\,{\text{d}}r 
  = 
  \frac{4\pi\, \nu(r)\, r^2\,{\text{d}}r}{L}.
\end{equation} 
This can be accomplished by generating a random deviate $X$ and solving the
equation $X=L(r)/L$ for $r$, where $L(r)$ is the total luminosity
emitted within a sphere with radius $r$,
\begin{equation}
  L(r)
  =
  4\pi \int_0^r \nu(r')\, r'^2\, {\text{d}}r'.
\end{equation}
For a S\'ersic model with S\'ersic index $n$, the function
$L(r)$ behaves smoothly and does not diverge \citep[see figure~2
in][]{1991A&A...249...99C}. It can in principle be calculated exactly
in terms of the Meijer G function \citep{2002A&A...383..384M}, but in
{\sc{skirt}} it is evaluated numerically. We have checked our results
for $L(r)$ with the analytical values for $n=1$ and with the tabulated
values for $n=4$ \citep{1976AJ.....81..807Y}.

\bsp

\end{document}